\pdfoutput=1

\documentclass[11pt]{article}

\usepackage{amsmath}
\usepackage{amsfonts}
\usepackage{amssymb}
\usepackage{bm}

\usepackage{algorithm}
\usepackage{algorithmic}

\usepackage{times}
\usepackage{url}

\usepackage{graphicx}
\usepackage[caption=false]{subfig}

\usepackage{array}
\usepackage{multirow}
\usepackage{rotating}

\usepackage{appendix}

\usepackage{authblk}
\usepackage[round]{natbib}
\usepackage{hyperref}

\DeclareMathOperator{\E}{E}
\DeclareMathOperator{\var}{var}
\DeclareMathOperator{\cov}{cov}
\DeclareMathOperator{\tr}{tr}
\DeclareMathOperator{\Stress}{Stress}
\DeclareMathOperator{\MStress}{MStress}
\DeclareMathOperator{\Energy}{Energy}
\DeclareMathOperator{\MEnergy}{MEnergy}
\DeclareMathOperator{\shortest}{shortest\_paths}
\DeclareMathOperator{\MDSweights}{MDS\_weights}
\renewcommand{\vec}[1]{\boldsymbol{\mathbf{#1}}}
\newcommand{\bo}{\mathbf}
\newcommand{\R}{\mathbb{R}}
\newcommand{\cost}{\mathcal{C}}

\begin{document}

\title{A Regularized Graph Layout Framework for Dynamic Network Visualization}

\renewcommand\Authfont{\small}
\renewcommand\Affilfont{\small}
\setlength{\affilsep}{3pt}
\author[1]{Kevin S.~Xu}
\author[2]{Mark Kliger}
\author[1]{Alfred O.~Hero III}
\affil[1]{EECS Department, University of Michigan, Ann Arbor, MI, USA \authorcr
\url{xukevin@umich.edu}, \url{hero@umich.edu}}
\affil[2]{Omek Interactive, Israel,
\url{mark.kliger@gmail.com}}

\maketitle

\begin{abstract}
Many real-world networks, including social and information networks, 
are dynamic structures that evolve over time. 
Such dynamic networks are typically visualized using a sequence of static 
graph layouts. 
In addition to providing a visual representation of the network structure at 
each time step, the sequence should preserve the \emph{mental map} between 
layouts of consecutive time steps to allow a human to interpret the temporal 
evolution of the network. 
In this paper, we propose a framework for dynamic network visualization 
in the \emph{on-line} setting where only present and past graph snapshots 
are available to create the present layout. 
The proposed framework creates \emph{regularized graph layouts} by 
augmenting the cost function of a static graph layout algorithm with a 
\emph{grouping penalty}, 
which discourages nodes from deviating too far from other nodes belonging to 
the same group, and a \emph{temporal penalty}, which discourages large 
node movements between consecutive time steps. 
The penalties increase the stability of the layout sequence, thus 
preserving the mental map. 
We introduce two dynamic layout algorithms within the proposed framework, 
namely dynamic multidimensional scaling (DMDS) and dynamic graph 
Laplacian layout (DGLL). 
We apply these algorithms on several data sets to illustrate the 
importance of both grouping and temporal regularization 
for producing interpretable visualizations of dynamic networks.
\end{abstract}

\section{Introduction}
The study of networks has emerged as a topic of great interest in recent 
years, with applications in the social, computer, and life sciences, 
among others. 
Dynamic networks are of particular 
interest because networks observed in the real world often evolve over time 
due to the creation of new nodes and edges and the removal of old nodes 
and edges \citep{KossinetsScience2006,LeskovecTKDD2007}. 
Many developments have been made in mining dynamic networks, 
including finding low-rank approximations \citep{SunSDM2007,TongKDD2008} 
and the detection of clusters or communities and how they evolve 
over time \citep{ChiTKDD2009,MuchaScience2010,Xu2013}. 
However, the closely related task of visualizing dynamic networks has 
remained an open problem that has attracted attention from sociologists 
\citep{Moody2005,Bender-deMoll2006,Leydesdorff2008} and the graph 
drawing community
\citep{Brandes1997,Branke2001,BrandesIV2003,Erten2004,Brandes2007,
Frishman2008,Brandes2011,Brandes2012} among others. 
Visualization is an important tool that can provide 
insights and intuition about networks that cannot be conveyed by summary 
statistics alone. 

Visualizing static networks is a challenge in itself. 
Static networks are typically represented by graphs, which have no 
natural representation in a Euclidean space. 
Many graph layout algorithms have been developed to create 
aesthetically pleasing $2$-D representations of graphs 
\citep{DiBattista1999,Herman2000}. 
Common layout methods for general graphs include 
force-directed layout \citep{Kamada1989,Fruchterman1991}, multidimensional 
scaling (MDS) \citep{DeLeeuw1980,Gansner2004,Borg2005} and graph Laplacian 
layout (GLL), also known as spectral layout 
\citep{Hall1970,Koren2005}. 

Dynamic networks are typically represented by a time-indexed sequence of 
graph snapshots; thus visualizing dynamic networks in $2$-D 
presents an additional challenge due to the temporal aspect 
\citep{Branke2001,Moody2005}. 
If one axis is used to represent time, then only a single axis remains to 
convey the topology of the network. 
While it is possible to identify certain trends from a $1$-D time plot 
created in this manner, it is a poor representation of the network structure. 
\citet{BrandesIV2003} presented a possible solution to this problem 
by creating a pseudo-$3$-D visualization that treats $2$-D layouts of each 
snapshot as layers in a stack. 
Unfortunately, the resulting visualization is difficult to interpret. 
The more conventional approach is to present an animated $2$-D layout that 
evolves over time to reflect the current snapshot 
\citep{Erten2004,Moody2005,Bender-deMoll2006,Brandes2007,Frishman2008}. 
A challenge with this approach is to preserve the \emph{mental map} 
\citep{Misue1995} between graph snapshots so that the transition between 
frames in the animation can 
be easily interpreted by a human viewer. 
In particular, it is undesirable for a large number of nodes to drastically 
change positions between frames, which may cause the viewer to lose 
reference of the previous layout. 

Some of the early work on dynamic network visualization simply involved 
creating interpolated transition layouts 
\citep{Moody2005,Bender-deMoll2006}. 
While interpolation does make an animation more aesthetically pleasing, it 
does not constrain the successive layouts in any way to make them more 
interpretable. 
In many real networks, individual snapshots have high variance, so creating 
a layout for each snapshot using a static graph layout method could result 
in large node movements between time steps. 
Often, this is not due to a failure of the static graph layout algorithm 
but simply a consequence of the cost function it is attempting to optimize, 
which does not consider any other snapshots. 

When dealing with dynamic networks, better performance can be obtained by 
using \emph{regularized} methods that consider both current and past 
snapshots. 
Such an approach has been used to develop regularized clustering algorithms 
for dynamic networks, also known as evolutionary clustering 
algorithms \citep{ChiTKDD2009,MuchaScience2010,Xu2013}, that 
outperform traditional static clustering algorithms in the dynamic setting. 
In the context of dynamic network visualization, 
regularization encourages layouts to be stable over time, thus 
preserving the mental map between snapshots. 
As such, many methods for dynamic graph drawing either implicitly or 
explicitly employ regularization in the form of a dynamic stability penalty 
to discourage large node movements 
\citep{Brandes1997,Branke2001,Erten2004,Baur2008,Frishman2008,Brandes2011,
Brandes2012}. 
The concept of regularization has also been employed in many problems in 
statistics and machine learning, including ridge regression 
\citep{Hoerl1970}, the LASSO \citep{TibshiraniJRSSB1996}, and penalized matrix 
decomposition \citep{Witten2009}. 
It is often used to introduce additional information or constraints and 
to give preference to solutions with certain desirable 
properties such as sparsity, smoothness, or in this paper, dynamic 
stability in order to preserve the mental map. 

We introduce a framework for dynamic network visualization 
using \emph{regularized graph layouts}. 
The framework is designed to generate layouts in the \emph{on-line} setting 
using only present and past snapshots. 
It involves optimizing a modified cost function that augments 
the cost function of a static graph layout algorithm with two penalties:
\begin{enumerate}
	\item A \emph{grouping penalty}, which discourages nodes from deviating 
		too far from other nodes belonging to the same group.
	\item A \emph{temporal penalty}, which discourages nodes from deviating 
		too far from their previous positions.
\end{enumerate}
Groups could correspond to a priori knowledge, 
such as participant affiliations in social networks. 
If no a priori group knowledge is available, groups can be learned from 
the network using, for example, the aforementioned evolutionary clustering 
algorithms. 
The grouping penalty keeps group members close together in the sequence of 
layouts, which helps to preserve the mental map because nodes belonging to 
the same group often evolve over time in a similar fashion. 
The temporal penalty helps to preserve the mental map by 
discouraging large node movements that may cause a human to lose reference 
of previous node positions, such as multiple nodes moving unnecessarily 
from one side of the layout to the opposite side. 
Within the proposed framework, 
we develop two dynamic layout algorithms, \emph{dynamic multidimensional 
scaling} (DMDS) and \emph{dynamic graph Laplacian layout} (DGLL).

To the best of our knowledge, this is the first framework for dynamic network 
visualization that incorporates \emph{both grouping and temporal 
regularization}\footnote{A preliminary version of this work can be found in 
\citep{XuMLG2011}.}. 
The methods for grouping regularization in DMDS 
and temporal regularization in DGLL used in this paper are novel. 
Temporal regularization in dynamic graph layouts has been employed in 
previous work 
\citep{Brandes1997,Branke2001,Erten2004,Baur2008,Frishman2008,Brandes2011,
Brandes2012}. 
Grouping regulariation 
has also appeared in previous work \citep{Wang1995,Eades2000,CostaICASSP2005}, 
but only in the static graph setting, not in the dynamic setting we consider. 
We apply the proposed DMDS and DGLL algorithms on a selection of dynamic 
network 
data sets to demonstrate the importance of both grouping and temporal 
regularization in creating interpretable visualizations. 

\section{Background}
We begin by specifying the notation we shall use in this paper. 
Time-indexed quantities are indicated using square brackets, e.g.~$X[t]$. 
We represent a dynamic network by a discrete-time sequence of graph snapshots. 
Each snapshot is represented by a graph adjacency matrix $W[t]$ where 
$w_{ij}[t]$ denotes the weight of the edge between nodes $i$ and $j$ at time 
$t$ (chosen to be $1$ for unweighted graphs), and $w_{ij}[t]=0$ if no edge is 
present. 
We assume all graphs are undirected, so that $w_{ij}[t] = w_{ji}[t]$. 
For simplicity of notation, we typically drop the time index for all 
quantities at time step $t$, i.e.~$W$ is assumed to denote $W[t]$. 

We refer to a graph layout by a matrix $X \in \R^{n \times s}$, where 
$n$ denotes the number of nodes present at time $t$, and each 
row $\bo x_{(i)}$ corresponds to the $s$-dimensional position of node $i$. 
We are most interested in $2$-D visualization ($s=2$), although 
the proposed methods can also be applied to other values of $s$. 
The $i$th column of $X$ is denoted by $\bo x_i$, and the individual entries 
by $x_{ij}$. 
The superscript in $\bo x_i^{(h)}$ is used to denote the value of $\bo x_i$ 
at iteration $h$ of an algorithm. 
The norm operator $\|\cdot\|$ refers to the $l_2$-norm, and 
$\tr(\cdot)$ denotes the matrix trace operator. 
We denote the all ones column vector by $\bo 1$. 

We now summarize the static graph layout methods of multidimensional scaling 
and graph Laplacian layout, which operate on a single graph snapshot. 
We develop regularized versions of these methods for dynamic networks in 
Section \ref{sec:Regularized}.

\subsection{Multidimensional scaling}
\label{sec:MDS}
Multidimensional scaling (MDS) is a family of statistical methods that aim 
to find an optimal layout $X \in \R^{n \times s}$ such that the distance 
between $\bo x_{(i)}$ and $\bo x_{(j)}$ for all $i \neq j$ is as close as 
possible 
to a desired distance $\delta_{ij}$. 
There are a variety of different cost functions and associated algorithms 
for MDS; we refer interested readers to \citet{Borg2005}. 
Here we describe the cost function known as stress and its associated 
majorization algorithm. 
The stress of a layout $X$ is given by 
\begin{equation}
	\label{eq:Stress}
	\Stress(X) = \frac{1}{2}\sum_{i=1}^n \sum_{j=1}^n v_{ij} 
		\left(\delta_{ij} - \|\bo x_{(i)} - \bo x_{(j)}\|\right)^2,
\end{equation}
where $v_{ij} \geq 0$ denotes the weight or importance of maintaining the 
desired distance $\delta_{ij}$. 
We refer to the matrix $V = [v_{ij}]$ as the \emph{MDS weight} matrix to 
avoid confusion with the graph adjacency matrix $W$, which is also sometimes 
referred to as a weight matrix. 
In order to use stress MDS for graph layout, the graph adjacency matrix 
$W$ is first converted into a desired distance matrix 
$\Delta = [\delta_{ij}]$. 
This is done by defining a distance metric over the graph and calculating 
distances between all pairs of nodes. 
The distance between two nodes $i$ and $j$ is typically taken to 
be the length of the shortest path between the nodes \citep{Gansner2004}. 
For weighted graphs, it is assumed that the edge weights denote 
dissimilarities; if the edge weights instead denote similarities, they must 
first be converted into dissimilarities before computing $\Delta$. 
The MDS weights $v_{ij}$ play a crucial role in the aesthetics of the layout. 
The commonly used Kamada-Kawai (KK) force-directed layout \citep{Kamada1989} 
is a special case of stress MDS where the weights are chosen to be 
$v_{ij} = \delta_{ij}^{-2}$ for $i \neq j$ and $v_{ii} = 0$.

The objective of stress MDS is to find a layout $X$ that minimizes 
\eqref{eq:Stress}. 
\eqref{eq:Stress} can be decomposed into 
\begin{equation}
	\label{eq:Stress_exp}
	\frac{1}{2}\sum_{i=1}^n \sum_{j=1}^n v_{ij} 
		\delta_{ij}^2 + \frac{1}{2}\sum_{i=1}^n \sum_{j=1}^n v_{ij} 
		\|\bo x_{(i)} - \bo x_{(j)}\|^2 - \sum_{i=1}^n \sum_{j=1}^n 
		v_{ij}\delta_{ij} \|\bo x_{(i)} - \bo x_{(j)}\|.
\end{equation}
Note that the first term of \eqref{eq:Stress_exp} is independent of $X$. 
The second term of \eqref{eq:Stress_exp} can be written as $\tr(X^T R X)$ 
where the $n \times n$ matrix $R$ is given by
\begin{equation}
	\label{eq:Def_R}
	r_{ij} = 
	\begin{cases}
		-v_{ij} & i \neq j,\\
		\sum_{k \neq i} v_{ik} & i = j.
	\end{cases}
\end{equation}
$\tr(X^T R X)$ is quadratic and convex in $X$ and is easily optimized. 

The third term of \eqref{eq:Stress_exp} cannot be written as a quadratic 
form. 
However, it can be optimized by an iterative majorization method known as 
``scaling by majorizing a complicated function'' (SMACOF) 
\citep{DeLeeuw1980,Borg2005}. 
This non-quadratic term is iteratively majorized, and the 
resulting upper bound for the stress is then optimized by differentiation. 
For a matrix $Z \in \R^{n \times s}$, define the matrix-valued function 
$S(Z)$ by 
\begin{equation}
	\label{eq:Def_S}
	s_{ij}(Z) = 
	\begin{cases}
		-v_{ij}\delta_{ij}/\|\bo z_{(i)} - \bo z_{(j)}\| & i \neq j,\\
		-\sum_{k \neq i} s_{ik}(Z) & i = j.
	\end{cases}
\end{equation}
Then, it is shown in \citep{Gansner2004,Borg2005} that 
\begin{equation*}
	-\tr(X^T S(Z) Z) \geq -\frac{1}{2}\sum_{i=1}^n \sum_{j=1}^n v_{ij} 
		\delta_{ij} \|\bo x_{(i)} - \bo x_{(j)}\|
\end{equation*}
so that an upper bound for the stress is 
\begin{equation}
	\label{eq:Stress_maj}
	\frac{1}{2} \sum_{i=1}^n \sum_{j=1}^n v_{ij} \delta_{ij}^2 + \tr(X^T R X) 
		- 2\tr(X^T S(Z) Z).
\end{equation}
By setting the derivative of \eqref{eq:Stress_maj} with respect to $X$ to 
$0$, the minimizer of the upper bound is found to be the solution to the 
equation $RX = S(Z)Z$. 

The algorithm for optimizing stress is iterative. 
Let $X^{(0)}$ denote an initial layout. 
Then at each iteration $h$, solve 
\begin{equation}
	\label{eq:MDS_linear_eq}
	R \, \bo x_a^{(h)} = S\left(X^{(h-1)}\right) \bo x_a^{(h-1)}
\end{equation}
for $\bo x_a^{(h)}$ for each $a = 1, \ldots, s$. 
\eqref{eq:MDS_linear_eq} can be solved using a standard 
linear equation solver. 
Note that $R$ is rank-deficient; this is a consequence of the stress function 
being translation-invariant \citep{Gansner2004}. 
The translation-invariance can be removed by fixing the location of one point, 
e.g.~by setting $\bo x_{(1)} = 0$, removing the first row and column of 
$R$, and removing the first row of $S\big(X^{(h-1)}\big) X^{(h-1)}$ 
\citep{Gansner2004}. 
\eqref{eq:MDS_linear_eq} can then be solved efficiently using Cholesky 
factorization. 
The iteration can be terminated when 
\begin{equation}
	\label{eq:Stress_convergence}
	\frac{\Stress\big(X^{(h-1)}\big) - \Stress\big(X^{(h)}\big)} 
		{\Stress\big(X^{(h-1)}\big)} < \epsilon,
\end{equation}
where $\epsilon$ is the convergence tolerance of the iterative process. 

\subsection{Graph Laplacian layout}
\label{sec:GLL}
Graph Laplacian layout (GLL) methods optimize a quadratic function associated 
with the graph Laplacian matrix \citep{Koren2005}, which we call the 
GLL energy. 
The graph Laplacian is obtained from the adjacency matrix 
by $L = D-W$, where $D$ is the diagonal matrix of node degrees defined by 
$d_{ii} = \sum_{j=1}^{n} w_{ij}$. 
For weighted graphs, GLL assumes that the weights correspond to similarities 
between nodes, rather than dissimilarities as in MDS. 
GLL is also referred to as ``spectral layout'' because the optimal solution 
involves the eigenvectors of the Laplacian, as we will show. 
The GLL energy function is defined by
\begin{equation}
	\label{eq:Energy}
	\Energy(X) = \frac{1}{2} \sum_{i=1}^n \sum_{j=1}^n w_{ij} \|\bo x_{(i)} 
		- \bo x_{(j)}\|^2.
\end{equation}
It is easily shown that $\Energy(X) = \tr(X^T L X)$. 
The GLL problem can be formulated as \citep{Hall1970,Koren2005}: 
\begin{align}
	\label{eq:GLL_obj}
	\min_X \quad &\tr(X^T L X) \\
	\label{eq:GLL_norm_cons}
	\text{subject to} \quad &X^T X = nI\\
	\label{eq:GLL_mean_cons}
	&X^T \bo 1 = \bo 0.
\end{align}
From \eqref{eq:Energy}, it can be seen that minimizing the GLL energy 
function aims to make edge lengths short by placing nodes connected by 
heavy edges close together. 
\eqref{eq:GLL_mean_cons} removes the trivial solution $\bo x_a = \bo 1$, 
which places all nodes at the same location in one dimension. 
It can also be viewed as removing a degree of freedom in the layout due 
to translation invariance \citep{BelkinNC2003} by setting the mean of 
$\bo x_a$ to $0$ for all $a$. 
Since $\bo x_a$ has zero-mean, $(\bo x_a^T \bo x_a)/n$ corresponds to the 
variance or scatter of the layout in dimension $a$. 
Thus \eqref{eq:GLL_norm_cons} constrains the layout to have unit 
variance in each dimension and zero covariance between 
dimensions so that each dimension of the layout provides as much 
additional information as possible. 
Moreover, one can see that \eqref{eq:GLL_norm_cons} differs slightly from 
the usual constraint $X^T X = I$ \citep{BelkinNC2003,Koren2005}, 
which constrains the layout to have variance $1/n$ in each dimension.
In the dynamic network setting where $n$ can vary over time, this is 
undesirable because the layout would change scale between time steps if the 
number of nodes changes. 

By a generalization of the Rayleigh-Ritz theorem \citep{Lutkepohl1997}, an 
optimal solution to the GLL problem is given by 
$X^* = \sqrt{n} [\bo v_2, \bo v_3, \ldots, 
\bo v_{s+1}]$, where $\bo v_i$ denotes the eigenvector corresponding to the 
$i$th smallest eigenvalue of $L$. 
Note that $\bo v_1 = (1/\sqrt{n})\bo 1$ is excluded because it violates the 
zero-mean constraint \eqref{eq:GLL_mean_cons}. 
Using the property that $\tr(ABC) = \tr(CAB)$, the cost function 
\eqref{eq:GLL_obj} is easily shown to be invariant to rotation and 
reflection, so $X^* R$ is also an optimal solution for any $R^T R = I$. 

In practice, it has been found that using degree-normalized eigenvectors 
often results in more aesthetically pleasing layouts 
\citep{BelkinNC2003,Koren2005}. 
The degree-normalized layout problem differs only in that the dot product 
in each of the constraints is replaced with the degree-weighted dot product, 
resulting in the following optimization problem:
\begin{equation*}
	\begin{split}
		\min_{X} \quad &\tr(X^T L X) \\
		\text{subject to} \quad &\tr(X^T D X) = \tr(D)I\\
		&X^T D \bo 1 = \bo 0.
	\end{split}
\end{equation*}
The optimal solution is given by $X^* = \sqrt{\tr(D)} \, [\bo u_2, \bo u_3, 
\ldots, \bo u_{s+1}]$ or any rotation or reflection of $X^*$, 
where $\bo u_i$ denotes the generalized eigenvector 
corresponding to the $i$th smallest generalized eigenvalue of $(L,D)$. 
Again, $\bo u_1 = \big(1/\sqrt{\tr(D)}\big)\bo 1$ is excluded because 
it violates the zero-mean constraint. 
A discussion on the merits of the degree normalization can be found in 
\citep{Koren2005}. 

\section{Regularized layout methods}
\label{sec:Regularized}

\subsection{Regularization framework}
\label{sec:Framework}
The aforementioned static layout methods can be applied snapshot-by-snapshot 
to create a visualization of a dynamic network; however, the resulting 
visualization is often difficult to interpret, especially if there are large 
node movements between time steps. 
We propose a regularized layout framework that uses a 
modified cost function, defined by
\begin{equation}
	\label{eq:Modified_cost}
	\cost_{\text{modified}} = \cost_{\text{static}} 
		+ \alpha\cost_{\text{grouping}} + \beta\cost_{\text{temporal}}.
\end{equation}
The \emph{static cost} $\cost_{\text{static}}$ corresponds to the cost 
function optimized by the static layout algorithm. 
For example, in MDS, it is the stress function defined in \eqref{eq:Stress}, 
and in GLL, it is the energy defined in \eqref{eq:Energy}. 
The \emph{grouping cost} $\cost_{\text{grouping}}$ is chosen to discourage 
nodes from 
deviating too far from other group members; $\alpha$ controls the importance 
of the grouping cost, so we refer to $\alpha\cost_{\text{grouping}}$ as the 
\emph{grouping penalty}. 
Similarly, the \emph{temporal cost} $\cost_{\text{temporal}}$ is chosen to 
discourage 
nodes from deviating too far from their previous positions; $\beta$ controls 
the importance of the temporal cost, so we refer to 
$\beta\cost_{\text{temporal}}$ as the \emph{temporal penalty}. 
We propose quadratic forms for these penalties, similar to 
ridge regression \citep{Hoerl1970}. 

Let $k$ denote the number of groups. 
Define the group membership by an $n \times k$ matrix $C$ where 
\begin{equation*}
	c_{il} = 
	\begin{cases}
		1 & \text{node $i$ is in group $l$ at time step $t$,}\\
		0 & \text{otherwise.}
	\end{cases}
\end{equation*}
We introduce grouping regularization by adding group representatives, 
which also get mapped to an $s$-dimensional position, stored in the matrix 
$Y \in \R^{k \times s}$. 
The proposed grouping cost is given by 
\begin{equation}
	\label{eq:Group_cost}
	\cost_{\text{grouping}}(X,Y) = \sum_{l=1}^k \sum_{i=1}^n c_{il} 
		\|\bo x_{(i)} - \bo y_{(l)}\|^2,
\end{equation}
where $\bo y_{(l)}$ denotes the position of the $l$th representative. 
Notice that the grouping cost is decreased by moving $\bo y_{(l)}$ and 
$\bo x_{(i)}$ towards each other if node $i$ is in group $l$. 
As a result, nodes belonging to the same group will be placed closer 
together than in a layout without grouping regularization. 
Notice also that we do not require knowledge of the group membership of 
every node. 
Nodes with unknown group memberships correspond to all-zero rows in $C$ 
and are not subject to any grouping penalty.

We introduce temporal regularization on nodes present at both time steps $t$ 
and $t-1$ by discouraging node positions from deviating significantly from 
their previous positions. 
This idea is often referred to in the graph drawing literature 
as maintaining \emph{dynamic stability} of the layouts and is often used to 
achieve the goal of preserving the mental map. 
Define the diagonal matrix $E$ by 
\begin{equation*}
	e_{ii} = 
	\begin{cases}
		1 & \text{node $i$ was present at time step $t-1$,}\\
		0 & \text{otherwise.}
	\end{cases}
\end{equation*}
The proposed temporal cost is then given by
\begin{equation}
	\label{eq:Temp_cost}
	\cost_{\text{temporal}}(X,X[t-1]) = \sum_{i=1}^n e_{ii} \|\bo x_{(i)} 
		- \bo x_{(i)}[t-1]\|^2.
\end{equation}
The temporal cost is decreased by moving $\bo x_{(i)}$ towards 
$\bo x_{(i)}[t-1]$, but unlike in the grouping cost, $\bo x_{(i)}[t-1]$ is 
fixed because it was assigned at the previous time step. 
Thus the previous node position acts as an anchor for the current node 
position. 

We note that the temporal cost measures only the stability of the layouts over 
time and is not necessarily a measure of goodness-of-fit with regard to the 
dynamic network. 
For example, if there is a sudden change in the network topology, 
an extremely low temporal cost may be undesirable because it could prevent the 
layout from adequately adapting to reflect this change. 
Thus one must consider the trade-off of adaptation rate versus 
stability when choosing the temporal regularization parameter. 

Next we demonstrate how the grouping and temporal 
penalties can be introduced into MDS and GLL as examples of the proposed 
regularization framework. 

\subsection{Dynamic multidimensional scaling}
\label{sec:DMDS}
The dynamic multidimensional scaling (DMDS) modified cost is given by the 
modified stress function 
\begin{equation}
	\label{eq:Reg_stress}
	\begin{split}
		\MStress(X&,Y) = \frac{1}{2} \sum_{i=1}^n \sum_{j=1}^n v_{ij} 
			\big(\delta_{ij} - \|\bo x_{(i)} - \bo x_{(j)}\|\big)^2 \\
		&+ \alpha \sum_{l=1}^k \sum_{i=1}^n c_{il} 
			\|\bo x_{(i)} - \bo y_{(k)}\|^2 + \beta \sum_{i=1}^n e_{ii} 
			\|\bo x_{(i)} - \bo x_{(i)}[t-1]\|^2.
	\end{split}
\end{equation}
The first term of \eqref{eq:Reg_stress} is the usual MDS stress function, 
while the second term corresponds to the grouping penalty, and the third
term corresponds to the temporal penalty. 
The constants $\alpha$ and $\beta$ are the grouping and temporal 
regularization parameters, respectively. 

To optimize \eqref{eq:Reg_stress}, we begin by re-writing the first two terms 
into a single term. 
Define the augmented MDS weight matrix by 
\begin{equation}
	\label{eq:Def_tilde_V}
	\tilde{V} = 
	\begin{bmatrix}
		V & \alpha C \\
		\alpha C^T & 0
	\end{bmatrix},
\end{equation}
where the zero corresponds to an appropriately sized all-zero matrix. 
Similarly, define the $(n+k) \times (n+k)$ augmented desired distance matrix 
$\tilde{\Delta}$ by filling the added rows and columns with zeros, i.e.~
\begin{equation}
	\label{eq:Def_tilde_Delta}
	\tilde{\Delta} = 
	\begin{bmatrix}
		\Delta & 0 \\
		0 & 0
	\end{bmatrix}
\end{equation} 
Let 
\begin{equation}
	\label{eq:Def_tilde_X}
	\tilde{X} = 
	\begin{bmatrix}
		X \\
		Y
	\end{bmatrix}
\end{equation}
denote the positions of the both the nodes and the group representatives.
Then, the first two terms of \eqref{eq:Reg_stress} can be written as 
\begin{equation*}
	\frac{1}{2} \sum_{i=1}^{n+k} \sum_{j=1}^{n+k} \tilde{v}_{ij} 
		\left(\tilde{\delta}_{ij} - \|\tilde{\bo x}_{(i)} 
		- \tilde{\bo x}_{(j)}\|\right)^2,
\end{equation*}
which has the same form as the usual stress defined in \eqref{eq:Stress}. 
The third term in \eqref{eq:Reg_stress} can be written as a quadratic 
function of $\tilde{X}$, namely
\begin{equation*}
	\beta\left[\tr\left(\tilde X^T \tilde E \tilde X\right) - 2\tr\left(
		\tilde X^T \tilde E \tilde X[t-1]\right) + \tr\left(\tilde 
		X^T[t-1] \tilde E \tilde X[t-1]\right)\right],
\end{equation*}
where the $(n+k) \times (n+k)$ matrix $\tilde{E}$ and the $(n+k) \times s$ 
matrix $\tilde{X}[t-1]$ are constructed by zero-filling as in the 
definition of $\tilde{\Delta}$.

Following the derivation in Section \ref{sec:MDS}, for any $(n+k) \times s$ 
matrix $Z$, \eqref{eq:Reg_stress} can be majorized by 
\begin{equation}
	\label{eq:Reg_majorized}
	\begin{split}
		\frac{1}{2} \sum_{i=1}^{n+k} \sum_{j=1}^{n+k} \, \tilde v_{ij} 
			\tilde \delta_{ij}^2 + \tr&(\tilde X^T \tilde R \tilde X) 
			- 2\tr(\tilde X^T \tilde S(Z) Z) 
			+ \beta\Big[\tr(\tilde X^T \tilde E \tilde X) \\
		&- 2\tr(\tilde X^T \tilde E \tilde X[t-1]) + \tr(\tilde X^T[t-1] 
			\tilde E \tilde X[t-1]) \Big],
	\end{split}
\end{equation}
where $\tilde R$ and $\tilde S$ are defined by substituting the augmented 
matrices $\tilde{V}$ and $\tilde{\Delta}$ for $V$ and $\Delta$, 
respectively, in \eqref{eq:Def_R} and \eqref{eq:Def_S}. 
\eqref{eq:Reg_majorized} is quadratic and convex in $X$ so the minimizer 
of the upper bound 
is found by setting the derivative of \eqref{eq:Reg_majorized} to 0, 
resulting in the equation 
\begin{equation*}
	\big(\tilde R + \beta \tilde E\big)\tilde X = \tilde S\big(Z\big) Z 
		+ \beta \tilde E \tilde X[t-1].
\end{equation*}
This can again be solved sequentially over each dimension. 
As in Section \ref{sec:MDS}, we solve this iteratively using the previous 
iteration as the majorizer, i.e. at iteration $h$, solve 
\begin{equation}
	\label{eq:DMDS_linear_eq}
	\big(\tilde R + \beta \tilde E\big)\tilde{\bo x}_a^{(h)} 
		= \tilde S\left(\tilde X^{(h-1)}\right) \tilde{\bo x}_a^{(h-1)} 
		+ \beta \tilde E \tilde{\bo x}_a[t-1]. 
\end{equation}
for $\tilde{\bo x}_a^{(h)}$ for each $a = 1, \ldots, s$. 
The process is iterated until the convergence criterion 
\eqref{eq:Stress_convergence} is attained.
The first iterate can be taken to be simply the previous layout 
$\tilde{\bo x}_a[t-1]$. 
Unlike in ordinary MDS, the system of linear 
equations in \eqref{eq:DMDS_linear_eq} has a unique solution provided that 
at least a single node was present at time step $t-1$, because 
$\tilde R+\beta \tilde E$ has full rank in this case. 

\begin{figure}[t]
	\begin{algorithmic}[1]
	\FOR {$t=1,2,\ldots$}
		\STATE $\Delta \leftarrow \shortest(W)$
		\STATE $V \leftarrow \MDSweights(\Delta)$
		\STATE Construct $\tilde{V}$ and $\tilde{\Delta}$ using 
			\eqref{eq:Def_tilde_V} and \eqref{eq:Def_tilde_Delta}, 
			respectively
		\STATE Construct $\tilde{R}$ by substituting $\tilde{V}$ for $V$ in 
			\eqref{eq:Def_R}
		\STATE $h \leftarrow 0$
		\STATE $\tilde{X}^{(0)} \leftarrow \tilde{X}[t-1]$
		\REPEAT
			\STATE $h \leftarrow h+1$
			\STATE Construct $\tilde{S}\big(\tilde{X}^{(h-1)}\big)$ by 
				substituting $\tilde{V}$, $\tilde{\Delta}$, and 
				$\tilde{X}^{(h-1)}$ for $V$, $\Delta$, and $Z$, 
				respectively, in \eqref{eq:Def_S}
			\FOR {$a = 1,\ldots,s$}
				\STATE Solve $\big(\tilde R + \beta \tilde E\big) 
					\tilde{\bo x}_a^{(h)} = \tilde S\big(\tilde X^{(h-1)} 
					\big) \tilde{\bo x}_a^{(h-1)} + \beta \tilde E 
					\tilde{\bo x}_a[t-1]$ for $\tilde{\bo x}_a^{(h)}$
			\ENDFOR
		\UNTIL {$\big[\MStress\big(\tilde{X}^{(h-1)}\big) - \MStress\big(
			\tilde{X}^{(h)}\big)\big] / \MStress\big(\tilde{X}^{(h-1)}
			\big) < \epsilon$}
		\STATE $\tilde{X} \leftarrow \tilde{X}^{(h)}$
	\ENDFOR
	\end{algorithmic}
	\caption{Pseudocode for the DMDS algorithm. The function 
		$\shortest(\cdot)$ computes the matrix of shortest paths between 
		all pairs of 
		nodes, and $\MDSweights(\cdot)$ computes the MDS weight matrix.}
	\label{fig:DMDS_alg}
\end{figure}

Pseudocode for the DMDS algorithm for $t=1,2,\ldots$ is shown in 
Fig.~\ref{fig:DMDS_alg}. 
\eqref{eq:DMDS_linear_eq} can be solved by performing a Cholesky 
factorization on $(\tilde{R} + \beta \tilde{E})$ followed by back 
substitution. 
At the initial time step ($t=0$), there are no previous node positions to 
initialize with, so a random initialization is used. 
Also, the position of one node should be fixed before solving 
\eqref{eq:DMDS_linear_eq} due to the translation-invariance discussed in 
Section \ref{sec:MDS}. 
The time complexity of the algorithm at all subsequent time steps 
is dominated by the $O(n^3)$ complexity of the Cholesky factorization, 
assuming $k \ll n$, but the factorization only needs 
to be computed at the initial iteration ($h=1$). 
All subsequent iterations require only matrix-vector products and back 
substitution and thus have $O(n^2)$ complexity. 

\subsection{Dynamic graph Laplacian layout}
\label{sec:DGLL}
The dynamic graph Laplacian layout (DGLL) modified cost is 
given by the modified energy function
\begin{equation}
	\label{eq:Reg_energy}
	\begin{split}
		\MEnergy(&X,Y) = \frac{1}{2} \sum_{i=1}^n \sum_{j=1}^n w_{ij} 
			\|\bo x_{(i)} - \bo x_{(j)}\|^2 \\
		&+ \alpha \sum_{l=1}^k \sum_{i=1}^n c_{il} \|\bo x_{(i)} 
			- \bo y_{(l)}\|^2 + \beta \sum_{i=1}^n e_{ii} 
			\|\bo x_{(i)} - \bo x_{(i)}[t-1]\|^2.
	\end{split}
\end{equation}
Like with DMDS, the first term of \eqref{eq:Reg_energy} is the usual GLL 
energy function, while the second term corresponds to the 
grouping penalty, and the third term corresponds to the temporal 
penalty. 
Again, the parameters $\alpha$ and $\beta$ correspond to the grouping and 
temporal regularization parameters, respectively. 

We first re-write \eqref{eq:Reg_energy} in a more compact form using 
the graph Laplacian. 
Define the augmented adjacency matrix by 
\begin{equation}
	\label{eq:Def_tilde_W}
	\tilde{W} = 
	\begin{bmatrix}
		W & \alpha C \\
		\alpha C^T & 0
	\end{bmatrix}.
\end{equation}
Notice that the group representatives have been added as nodes to the graph, 
with edges between each node and its associated representative of weight 
$\alpha$. 
Define the augmented degree matrix by $\tilde{D}$ by $\tilde d_{ii} 
= \sum_{j=1}^{n+k} \tilde{w}_{ij}$, and the augmented graph Laplacian by 
$\tilde{L} = \tilde{D} - \tilde{W}$.
The first two terms of \eqref{eq:Reg_energy} can thus be written as 
$\tr(\tilde{X}^T \tilde L \tilde{X})$, where $\tilde{X}$ is as 
defined in \eqref{eq:Def_tilde_X}. 
The third term of \eqref{eq:Reg_energy} can be written as 
\begin{equation}
	\label{eq:DGLL_temporal}
	\beta \left[\tr(\tilde{X}^T \tilde{E} \tilde{X}) - 2\tr(\tilde{X}^T 
	\tilde{E} \tilde{X}[t-1]) + \tr(\tilde{X}^T[t-1] \tilde{E} 
	\tilde{X}[t-1])\right], 
\end{equation}
where $\tilde{E}$ is zero-filled as described in Section \ref{sec:DMDS}. 
The final term in \eqref{eq:DGLL_temporal} is independent of 
$\tilde{X}$ and is henceforth dropped from the modified cost. 

We now consider the constraints, which differ depending on whether the 
layout is degree-normalized, as discussed in Section \ref{sec:GLL}. 
We derive the constraints for the degree-normalized layout; the equivalent 
constraints for the unnormalized layout can simply be obtained by replacing 
$\tilde D$ with the identity matrix in the derivation. 
First we note that, due to the temporal regularization, the optimal layout 
is no longer translation-invariant, so we can remove the zero-mean 
constraint. 
As a result, the variance and orthogonality constraints become more complicated 
because we need to subtract the mean. 
Denote the degree-weighted mean in dimension $a$ by 
\begin{equation*}
	\tilde{\mu}_a = \frac{1}{\sum_{i=1}^{n+k} \tilde{d}_{ii}} \sum_{i=1}^{n+k}
		\tilde{d_{ii}} \tilde{x}_{ia}. 
\end{equation*}
Then the degree-weighted covariance between the $a$th and $b$th dimensions is 
given by 
\begin{align*}
	\cov &(\tilde{\bo x}_a,\tilde{\bo x}_b) = \frac{1}{\sum_{i=1}^{n+k} 
		\tilde{d}_{ii}} \sum_{i=1}^{n+k} \tilde{d}_{ii} (\tilde{x}_{ia} 
		- \tilde{\mu}_a) (\tilde{x}_{ib} - \tilde{\mu}_b) \\
	&= \frac{1}{\sum_{i=1}^{n+k} \tilde{d}_{ii}} \sum_{i=1}^{n+k} 
		\tilde{d}_{ii} \tilde{x}_{ia} \tilde{x}_{ib} - \frac{1} 
		{\left(\sum_{i=1}^{n+k} \tilde{d}_{ii}\right)^2} \left(\sum_{i=1}^{n+k} 
		\tilde{d}_{ii} \tilde{x}_{ia}\right) \left(\sum_{i=1}^{n+k} 
		\tilde{d}_{ii} \tilde{x}_{ib}\right) \\
	&= \frac{\tilde{\bo x}_a^T M \tilde{\bo x}_b}{\tr(\tilde D)},
\end{align*}
where $M$ is the centering matrix defined by 
\begin{equation}
	\label{eq:Def_center}
	M = \tilde{D} - \frac{\tilde{D} \bo 1 \bo 1^T \tilde{D}}{\tr(\tilde{D})}.
\end{equation}

Combining the modified cost function with the modified constraints, the 
normalized DGLL problem is as follows:
\begin{align}
	\label{eq:DGLL_obj}
	\min_{\tilde{X}} \quad &\tr(\tilde{X}^T \tilde{L} 
		\tilde{X}) + \beta\left[\tr(\tilde{X}^T \tilde{E} 
		\tilde{X}) - 2\tilde{X}^T \tilde{E} 
		\tilde{X}[t-1])\right] \\
	\label{eq:DGLL_const}
	\text{subject to} \quad &\tr(\tilde{X}^T M \tilde{X}) 
		= \tr(\tilde{D}) I.
\end{align}
Again, the unnormalized problem can be obtained by replacing $\tilde{D}$ with 
the identity matrix in \eqref{eq:Def_center} and \eqref{eq:DGLL_const}. 
Note that \eqref{eq:DGLL_obj} contains a linear term 
in $\tilde{X}$. 
Hence the optimal solution cannot be obtained using scaled generalized 
eigenvectors as in the static GLL problem. 
\eqref{eq:DGLL_obj} can be solved using standard algorithms for 
constrained nonlinear optimization \citep{Bazaraa2006}. 
The cost function and constraints consist only of linear and quadratic 
terms, so the gradient and Hessian are easily computed in closed form 
(see Appendix \ref{sec:DGLL_2D}). 
Unfortunately, the problem is not convex due to the equality constraints; 
thus a good initialization is important. 
The natural choice is to initialize using the previous layout 
$\tilde{X}^{(0)} = \tilde{X}[t-1]$. 
To avoid getting stuck in poor local minima, one could use multiple 
restarts with random initialization. 

\begin{figure}[t]
	\begin{algorithmic}[1]
	\FOR {$t=1,2,\ldots$}
		\STATE Construct $\tilde{W}$ using \eqref{eq:Def_tilde_W} and its 
			corresponding Laplacian $\tilde{L} = \tilde{D} - \tilde{W}$
		\STATE Construct the centering matrix $M$ using \eqref{eq:Def_center}
		\STATE $\tilde{X}^{(0)} \leftarrow \tilde{X}[t-1]$
		\STATE Solve \eqref{eq:DGLL_obj} using the forms for $\nabla f$, 
			$g$, $H$, and $J$ in Appendix \ref{sec:DGLL_2D}
		\FOR [if multiple random restarts are necessary] 
			{$r=1 \to max\_restarts$}
			\STATE Randomly assign $\tilde{X}^{(0)}$
			\STATE Solve \eqref{eq:DGLL_obj} using the forms for 
				$\nabla f$, $g$, $H$, and $J$ in Appendix 
				\ref{sec:DGLL_2D}
		\ENDFOR
		\STATE $\tilde{X} \leftarrow$ best solution to \eqref{eq:DGLL_obj} 
			over all initializations
	\ENDFOR
	\end{algorithmic}
	\caption{Pseudocode for the DGLL algorithm.}
	\label{fig:DGLL_alg}
\end{figure}

Pseudocode for the DGLL algorithm for $t=1,2,\ldots$ is shown in 
Fig.~\ref{fig:DGLL_alg}. 
We use the interior-point algorithm of \citet{Byrd1999} to solve 
\eqref{eq:DGLL_obj}. 
We find in practice that random restarts are not necessary unless 
$\beta$ is extremely small because the temporal regularization penalizes
solutions that deviate too far from the previous layout. 
For other choices of $\beta$, we find that the interior-point algorithm 
indeed converges to the global minimum when initialized using the previous 
layout. 
The most time-consuming operation in solving \eqref{eq:DGLL_obj} consists 
of a Cholesky factorization, which must be updated at each iteration. 
At the initial time step ($t=0$), there are no previous node positions, 
and hence, no linear term in 
\eqref{eq:DGLL_obj}, so the layout is obtained using scaled 
generalized eigenvectors, as described in Section \ref{sec:GLL}. 
The time complexity at all subsequent time steps is dominated by 
the $O(n^3)$ complexity of the Cholesky factorization. 

\subsection{Discussion}
\label{sec:Discussion}
We chose to demonstrate the proposed framework with MDS and GLL; however, it 
is also applicable to other graph layout methods, such as the 
Fruchterman-Reingold method of force-directed layout \citep{Fruchterman1991}. 
Since the static cost functions of MDS and GLL encourage different 
appearances, the same is true of DMDS and DGLL. 
Ultimately, the decision of which type of layout to use depends on the 
type of network and user preferences. 
Kamada-Kawai MDS layouts are often preferred in $2$-D because they discourage 
nodes from overlapping due to the large MDS weights assigned to maintaining 
small desired distances. 
On the other hand, if a $1$-D layout is desired, so that the entire sequence 
can be plotted as a time series, node overlap is a lesser concern. 
For such applications, DGLL may be a better choice. 

Another decision that needs to be made by the user is the choice of the 
parameters $\alpha$ and $\beta$, which 
can be tuned as desired to create a meaningful animation. 
Unlike in supervised learning tasks such as classification, there is no 
ground truth in visualization so the selection of parameters in 
layout methods is typically done in an ad-hoc fashion. 
Furthermore, multiple layouts created by differing choices of parameters 
could be useful for visualizing different portions of the network or 
yielding different insights \citep{WittenCSDA2011}. 
This is particularly true of the grouping regularization parameter $\alpha$. 
When a high value of $\alpha$ is used, nodes belonging to the same group are 
placed much closer together than nodes belonging to different groups. 
The resulting visualization emphasizes node movements between groups (for 
nodes that change group between time steps) while 
sacrificing the quality of the node movements within groups. 
On the other hand, when a low value of $\alpha$ is used, node movements 
within groups are more clearly visible, but node movements between groups 
are more difficult to see. 
We explore the effect of changing parameters on the resulting animation in 
several experiments in Section \ref{sec:Experiments}.

Finally, we note that the proposed framework is designed for the on-line 
setting where only past and present graph snapshots are available. 
Hence the temporal cost \eqref{eq:Temp_cost} involves only node positions 
at times $t$ and $t-1$, not $t+1$. 
In the off-line setting where the entire sequence of snapshots is available 
in advance, one can obtain higher quality layout sequences using an off-line 
method as discussed in \citet{Brandes2012}. 
The proposed framework could easily be modified for the off-line 
setting if desired. 
The grouping cost \eqref{eq:Group_cost} would not need any modification, and 
the temporal cost \eqref{eq:Temp_cost} would simply become 
$\cost_{\text{temporal}}(X,X[t-1]) 
+ \cost_{\text{temporal}}(X,X[t+1])$ to discourage nodes from deviating from 
both their past and future positions. 
$\cost_{\text{modified}}$ would then be optimized over all times $t$ 
simultaneously rather than snapshot-by-snapshot.

\section{Related work}
The regularized graph layout framework proposed in this paper utilizes 
a \emph{grouping} 
penalty that places nodes belonging to the same group together and 
a \emph{temporal} 
penalty that places nodes near their positions at neighboring time steps. 
Node grouping and temporal stability in the context of graph layout have 
previously been studied independently of each other. 
We summarize relevant contributions to both of these areas.

\subsection{Node grouping}
Several grouping techniques for static graph layout have been previously been 
proposed. 
Given a partition of a graph into groups, \citet{Wang1995} propose a modified 
force-directed layout that considers three types of forces: intra-forces, 
inter-forces, and meta-forces. 
Intra-forces and inter-forces denote forces between nodes in the same group 
and nodes in different groups, respectively. 
Meta-forces correspond to forces between groups; nodes in the same group are 
subject to identical meta-forces. 
By decreasing the strength of inter-forces and increasing the strength of 
meta-forces, nodes belonging to the same group get positioned closer to each 
other in the layout. 

\citet{Eades2000} developed a system called DA-TU for visualizing groups in 
large static graphs. 
It also utilizes a modified force-directed layout with inter- and 
intra-forces. 
However, rather than using meta-forces, DA-TU adds a virtual node for 
each group with a virtual force between each virtual node and each node 
in its group. 
Notice that the virtual nodes are identical to the group representatives in 
our proposed framework; however, the use of virtual forces to achieve 
grouping regularization differs from the squared Euclidean distance grouping 
cost we propose. 
In addition, DA-TU was designed 
for visualizing static graphs at many scales rather than visualizing dynamic 
graphs, so it does not contain a temporal stability penalty. 

Grouping techniques have been applied in the field of supervised 
dimensionality reduction, which is very closely related to graph layout. 
The objective of dimensionality reduction (DR) is to find a mapping $\phi: 
\R^p \rightarrow \R^s, \, p>s$ from a high-dimensional space 
to a lower-dimensional space while preserving many of the characteristics 
of the data representation in the high-dimensional space \citep{Lee2007}. 
For example, MDS is a DR method that attempts to 
preserve pairwise distances between data points. 
In the supervised DR setting, one also has a priori knowledge of the 
grouping structure of the data. 
Supervised DR methods pose the additional constraint that data points within 
the same group should be closer together in the low-dimensional space than 
points in separate groups. 
Notice that this is the same grouping constraint we pose in our regularized 
layout framework. 

\citet{WittenCSDA2011} proposed a supervised version of 
MDS (SMDS) that optimizes the following cost function over $X$: 
\begin{equation}
	\label{eq:Witten_stress}
	\frac{1}{2} \sum_{i=1}^n \sum_{j=1}^n (\delta_{ij} 
		- \|\bo x_{(i)} - \bo x_{(j)}\|)^2 
		+ \alpha \sum_{i,j: y_j > y_i} 
		(y_j - y_i) \sum_{a=1}^s \left(\frac{\delta_{ij}}{\sqrt{s}} 
		- (x_{ja} - x_{ia})\right)^2
\end{equation}
where $y_i$ is an ordinal value denoting the group membership of data 
point $i$. 
Notice that the first term in \eqref{eq:Witten_stress} is the ordinary 
MDS stress with $v_{ij} = 1$ for all $i,j$, while the second term provides 
the grouping regularization. 
$\alpha$ controls the trade-off between the two terms. 
The key difference between the SMDS grouping penalty 
and the DMDS grouping penalty proposed in this 
paper is in the way groups are treated. 
SMDS assumes that groups are labeled with an ordinal 
value that allows them to be ranked, and the form of the grouping penalty in 
\eqref{eq:Witten_stress} does indeed tend to rank groups in $\R^s$ 
by encouraging $x_{ja} > x_{ia}, a = 1, \ldots, s$ for all $i,j$ such that 
$y_j > y_i$. 
On the other hand, our proposed grouping penalty treats group labels as 
categorical. 
It does not rank groups in $\R^s$ 
but simply pulls nodes belonging to the group together. 

Another related method for supervised DR is classification 
constrained dimensionality reduction (CCDR) 
\citep{CostaICASSP2005}, which is a supervised version of Laplacian 
eigenmaps \citep{BelkinNC2003}. 
CCDR optimizes the following cost function over $(X,Y)$: 
\begin{equation*}
	\frac{1}{2}\sum_{i=1}^n \sum_{j=1}^n \|\bo x_{(i)} - \bo x_{(j)}\|^2 
		+ \alpha \sum_{l=1}^k \sum_{i=1}^n c_{il} \|\bo x_{(i)}
		- \bo y_{(l)}\|^2.
\end{equation*}
Notice that this cost function is simply the sum of the GLL energy and 
the DGLL grouping penalty. 
Indeed, DGLL can be viewed as an extension of CCDR to time-varying data. 
The CCDR solution is given by the matrix of generalized eigenvectors 
$\tilde U = [\bo{\tilde u}_2, \ldots, \bo{\tilde u}_{s+1}]$ of 
$(\tilde L,\tilde D)$, where 
$\tilde{L}$ and $\tilde{D}$ are as defined in Section \ref{sec:DGLL}. 
Although the addition of the temporal regularization due to the anchoring 
presence of the previous layout $X[t-1]$ prevents the DGLL 
problem from being solved using generalized eigenvectors, it discourages 
large node movements between time steps in order to better preserve the 
mental map.

\subsection{Temporal stability}
\label{sec:Rel_layout}
There have been many previous studies on the problem of laying out 
dynamic networks while preserving stability between time snapshots. 
\citet{Moody2005} proposed to generate dynamic layouts 
using a static 
layout method such as Kamada-Kawai MDS and to initialize at each 
time step using the layout generated at the previous time step. 
The approach of \citet{Moody2005} is implemented in the social network 
visualization software SoNIA \citep{SoNIA}. 
Such an approach, however, does not allow one to explicitly control the 
stability of the layout sequence, as noted by \citet{Brandes2012}. 
We also find it to be insufficient 
at preventing drastic node movements over time in our experiments 
in Section \ref{sec:Experiments}. 

To enforce stability in layouts at consecutive time steps, \citet{Brandes1997} 
proposed a probabilistic framework for dynamic network layout where the 
objective is to choose the layout at a particular time step with maximum 
posterior probability given the previous layout. 
Similar to our proposed framework, the probabilistic framework is applicable 
to a wide class of graph layout methods. 
\citet{Brandes1997} proposed several different criteria for temporal stability. 
One such criterion is to demand stability of node positions in consecutive 
layouts. 
For this notion of stability, the authors model each node's current position 
by a spherical Gaussian distribution centered at the node's previous position.
Thus the posterior probability can be written as (up to a normalizing 
constant)
\begin{equation}
	\label{eq:BW_posterior}
	\exp\left\{-\left(\cost_{\text{static}} + \frac{\sum_{i=1}^n e_{ii} 
		\|\vec{x}_{(i)} - \vec{x}_{(i)}[t-1]\|^2}{2\sigma^2}\right)\right\},
\end{equation}
where $\sigma$ is a scaling parameter for the amplitude of node movements. 
Notice that by taking the logarithm of \eqref{eq:BW_posterior}, one obtains 
the same form as our proposed regularized framework \eqref{eq:Modified_cost}, 
excluding the grouping cost, with $\beta = 1/(2\sigma^2)$. 
The layout that maximizes the logarithm of \eqref{eq:BW_posterior} is the 
same layout that maximizes posterior probability; 
thus, under this notion of stability, there is an equivalence between the 
probabilistic framework of \citet{Brandes1997} and the temporal regularization 
framework proposed in this paper. 

Other methods for preserving temporal stability in dynamic layouts tend to 
be specific to a particular layout method. 
\citet{Baur2008} proposed a temporally regularized MDS algorithm 
that uses the following localized update rule at each 
iteration $h$ for each node $i$ at each time step $t$:
\begin{equation}
	\label{eq:Visone_update}
	x_{ia}^{(h)} = \frac{\tilde{x}_{ia}^{(h-1)} + \beta(e_{ii} x_{ia}[t-1] 
		+ f_{ii} x_{ia}[t+1])}{\sum_{j \neq i} v_{ij} + \beta(e_{ii}+f_{ii})},
\end{equation}
where 
\begin{equation*}
	\tilde{x}_{ia}^{(h-1)} = \sum_{j \neq i} v_{ij} \left(x_{ja}^{(h-1)} 
		+ \delta_{ij} \frac{x_{ia}^{(h-1)}-x_{ja}^{(h-1)}} 
		{\|\bo x_{(i)}^{(h-1)} - \bo x_{(j)}^{(h-1)}\|}\right),
\end{equation*}
and $F$ is the diagonal matrix defined by 
\begin{equation*}
	f_{ii} = 
	\begin{cases}
		1 & \text{node $i$ is present at time step $t+1$,}\\
		0 & \text{otherwise.}
	\end{cases}
\end{equation*}
This algorithm was used in \citet{Leydesdorff2008} 
for visualizing similarities in journal content over time. 
\eqref{eq:Visone_update} is an off-line update because it 
uses both the node positions at time steps $t-1$ and $t+1$ 
to compute the node position at time step $t$, whereas the methods we 
propose, including DMDS, are on-line methods that use only current and 
past data. 
It was shown in \citet{Baur2008} that the localized update of 
\eqref{eq:Visone_update} optimizes the sum of the $\MStress$ 
function in \eqref{eq:Reg_stress} over all $t$ 
with $k=0$, i.e.~without a grouping penalty. 
It is one of many possible ways to add temporal stability to layouts created 
by stress minimization in the off-line setting. 
\citet{Brandes2011} performed a quantitative comparison of different temporal 
stability penalties for MDS layouts, mostly for the off-line setting, 
including that of \citet{Baur2008}.  
Several of these penalties are implemented in the social network analysis and 
visualization software Visone \citep{Brandes2004,Visone}. 
\eqref{eq:Visone_update} can be modified into an on-line update by 
removing the terms involving $f_{ii}$; 
the on-line modification optimizes the $\MStress$ function in 
\eqref{eq:Reg_stress} at a single time step with $k=0$, i.e.~without a 
grouping penalty\footnote{This on-line modification is referred to in 
\citet{Brandes2011} as the APP (anchor to previous layout, initialized with 
previous layout) method.}. 
Hence the proposed DMDS layout method can 
be viewed as an on-line modification of the method of \citet{Baur2008} 
with the addition of a grouping penalty. 

When it comes to GLL, to the best of our knowledge, there is no prior work 
that explicitly enforces temporal stability. 
\citet{Brandes2007} suggested two approaches for interpolating between static 
spectral layouts, where eigenvectors are calculated using power iteration 
\citep{Trefethen1997} initialized with the layout at the previous time step, 
similar to the approach of \citet{Moody2005} for MDS. 
The first is to simply linearly interpolate node positions between layouts, 
and the second is to compute a spectral layout of a smoothed 
graph Laplacian matrix $\lambda L[t-1] + (1-\lambda)L[t]$. 
The second approach utilizes both current and past graph snapshots and 
should perform better than the first approach, but it also does not explicitly 
constrain the layout at time $t$ from 
deviating too far from the layout at time $t-1$. 
We henceforth refer to the second approach as the BFP method and use it as a 
baseline for evaluating the performance of our proposed DGLL algorithm. 

Other methods for laying out dynamic networks have also been proposed. 
TGRIP \citep{Erten2004} is a modified force-directed layout method 
with added edges between vertices present at multiple time steps. 
The user-selected weights of these added edges control the amount of 
temporal regularization in the layouts. 
The method of \citet{Frishman2008} is also a 
modified force-directed layout. 
It is an on-line method that uses pinning weights to 
previous node positions to achieve temporal regularization and a GPU-based 
implementation to reduce run-time. 
The emphasis in both methods is on improving scalability to deal with 
extremely large networks by coarsening graphs to compute an initial 
layout then applying local refinements to improve the quality of the 
layout. 
As a result, they are applicable to much larger networks than the 
$O(n^3)$ methods proposed in this paper.
However, these methods do not incorporate any sort of grouping regularization 
to discourage nodes from deviating too far from other nodes in the same 
group. 

\section{Experiments}
\label{sec:Experiments}
We demonstrate the proposed framework by applying DMDS and DGLL on a 
simulated data set and two real data sets. 
Several snapshots of the resulting visualizations are presented. 
The full, animated visualizations over all time steps can be found 
on the supporting website \citep{XuSuppWeb}. 

In the second experiment, we do not have a priori group knowledge. 
Hence we learn the groups using the AFFECT evolutionary 
spectral clustering algorithm \citep{Xu2013}, summarized in Appendix 
\ref{sec:AFFECT}. 
In the other two experiments, we do have a priori group knowledge. 
We compute layouts both using the known groups and the groups learned 
by clustering. 
We also compute layouts using several other methods as baselines for 
comparison. 
Since the proposed framework is designed for the on-line setting, we use 
only other on-line methods as baselines. 
DMDS is compared to static MDS initialized using the previous layout 
as in \citet{Moody2005} and the on-line modification of the method of 
\citet{Baur2008} discussed in Section \ref{sec:Rel_layout}, which we 
denote by ``stabilized MDS''. 
DGLL is compared to the CCDR method of \citet{CostaICASSP2005}, the BFP 
method of computing the spectral layout of a smoothed graph Laplacian matrix 
\citep{Brandes2007}, and the standard spectral GLL solution \citep{Koren2005}. 
Note that the proposed DMDS and DGLL methods are the only ones that utilize 
both grouping and temporal regularization; the other baselines either use 
only grouping regularization (CCDR), only temporal regularization 
(stabilized MDS), or neither. 

Summary statistics from the experiments are presented in Tables 
\ref{tab:Costs_DMDS} and \ref{tab:Costs_DGLL} for the MDS- and 
GLL-based methods, respectively, and are discussed in Sections 
\ref{sec:SBM}--\ref{sec:Reality}. 
The KK choice of MDS weights is used for all of the MDS-based methods, 
and degree-normalized layout is used for all of the GLL-based methods. 

We define three measures of layout quality: static cost, 
centroid cost, and temporal cost. 
The \emph{static cost} measures how well the current layout coordinates fit 
the current graph snapshot. 
It is the cost function that would be optimized by the static graph layout 
algorithm, either MDS or GLL. 
The static cost for the MDS-based methods is taken to be the static MDS 
stress defined in \eqref{eq:Stress}. 
The static cost for the GLL-based methods is the GLL energy defined in 
\eqref{eq:Energy}. 

The \emph{centroid cost} is the sum of squared distances between each node 
and its group centroid, which is 
also the cost function of the well-known method of k-means clustering. 
It is used to measure how close nodes are to members of their 
group\footnote{Note that we cannot simply use the grouping cost 
\eqref{eq:Group_cost} because it is not 
defined for methods that do not incorporate grouping regularization.}. 
When prior knowledge of the groups is available, we calculate the centroid 
cost with respect to the known groups, even for the layouts where groups 
are learned by clustering. 
When prior knowledge is not available, we calculate the centroid cost with 
respect to the learned groups. 

The \emph{temporal cost} \eqref{eq:Temp_cost} is the sum of squared distances 
between node positions in layouts at consecutive time steps. 
It is often used to quantify how well the mental map is 
preserved over time \citep{Brandes1997,Branke2001,Baur2008,
Frishman2008,Brandes2011,Brandes2012}. 
As mentioned in Section \ref{sec:Framework}, the temporal cost should only 
be interpreted as a measure of stability and not a measure of temporal 
goodness-of-fit. 

The costs displayed are appropriately normalized 
(either by the number of nodes 
or pairs of nodes, depending on the quantity) so they are 
comparable across different data sets. 
For the MDS-based methods, we also compare the number of iterations required 
for convergence to a tolerance of $\epsilon = 10^{-4}$. 
For the BFP method, the parameter $\lambda$ lies on a different scale from 
the parameters $\alpha$ and $\beta$ for the other methods. 
To ensure a fair comparison, we choose $\lambda$ to 
minimize $\cost_\text{static} + \alpha \cost_\text{centroid} + \beta 
\cost_\text{temporal}$, 
where $\alpha$ and $\beta$ are chosen to be the same parameters used for the 
other methods. 

\begin{sidewaystable}
	\setlength{\extrarowheight}{1pt}
	\centering
	\begin{tabular}{cccccc}
		\hline
		Experiment & Algorithm & MDS stress & Centroid cost 
			& Temporal cost & Iterations\\
		\hline
		\multirow{4}{*}{SBM} & DMDS (known) & $0.160 \pm 0.000$ 
			& $\bo{0.257 \pm 0.001}$ & $\bo{0.262 \pm 0.001}$ 
			& $\bo{45.6 \pm 0.3}$\\
		& DMDS (learned) & $0.160 \pm 0.000$ & $0.308 \pm 0.003$ 
			& $0.293 \pm 0.002$ & $46.9 \pm 0.4$\\
		& Stabilized MDS & $0.157 \pm 0.000$ & $0.434 \pm 0.003$ 
			& $0.340 \pm 0.002$ & $51.0 \pm 0.4$\\
		& Static MDS & $\bo{0.132 \pm 0.000}$ & $0.623 \pm 0.004$ 
			& $1.271 \pm 0.010$ & $112.9 \pm 1.0$\\
		\hline
		\multirow{3}{*}{Newcomb} & DMDS (learned) & $0.136$ & $\bo{0.656}$ 
			& $\bo{0.089}$ & $\bo{13.8}$\\
		& Stabilized MDS & $0.107$ & $1.275$ & $0.125$ & $16.9$\\
		& Static MDS & $\bo{0.065}$ & $1.611$ & $1.368$ & $68.5$\\
		\hline
		\multirow{4}{*}{MIT} & DMDS (known) & $0.154$ & $\bo{1.334}$ 
			& $\bo{0.207}$ & $\bo{37.7}$\\
		& DMDS (learned) & $0.154$ & $1.491$ & $0.241$ & $38.6$\\
		& Stabilized MDS & $0.142$ & $1.900$ & $0.290$ & $45.8$\\
		& Static MDS & $\bo{0.092}$ & $2.637$ & $3.384$ & $108.3$\\
		\hline
	\end{tabular}
	\caption{Mean costs of MDS-based layouts ($\pm$ standard 
		error for SBM simulation experiment). 
		The smallest quantity (within one standard error) in each 
		column for each experiment is bolded. 
		DMDS results using both a priori known groups (when available) and 
		groups learned by clustering are shown. 
		The regularizers present in DMDS lower both the centroid and 
		temporal costs at the expense of higher static cost.}
	\label{tab:Costs_DMDS}
\end{sidewaystable}

\begin{sidewaystable}
	\setlength{\extrarowheight}{1pt}
	\centering
	\begin{tabular}{ccccc}
		\hline
		Experiment & Algorithm & GLL energy & Centroid cost 
			& Temporal cost\\
		\hline
		\multirow{5}{*}{SBM} & DGLL (known) & $0.658 \pm 0.003$ 
			& $\bo{0.294 \pm 0.002}$ & $\bo{0.732 \pm 0.007}$\\
		& DGLL (learned) & $0.655 \pm 0.003$ & $0.403 \pm 0.005$ 
			& $0.858 \pm 0.009$\\
		& CCDR & $0.630 \pm 0.003$ & $0.414 \pm 0.003$ 
			& $2.678 \pm 0.025$\\
		& BFP & $0.636 \pm 0.003$ & $0.637 \pm 0.006$ 
			& $2.269 \pm 0.027$\\
		& Spectral & $\bo{0.605 \pm 0.003}$ & $0.945 \pm 0.007$ 
			& $3.179 \pm 0.022$\\
		\hline
		\multirow{4}{*}{Newcomb} & DGLL (learned) & $0.820$ & $1.325$ 
			& $\bo{0.379}$\\
		& CCDR & $0.786$ & $1.334$ & $1.352$\\
		& BFP & $0.783$ & $\bo{1.321}$ & $0.793$\\
		& Spectral & $\bo{0.761}$ & $1.373$ & $1.425$\\
		\hline
		\multirow{5}{*}{MIT} & DGLL (known) & $0.130$ & $\bo{1.228}$ 
			& $\bo{0.263}$\\
		& DGLL (learned) & $0.128$ & $1.357$ & $0.313$\\
		& CCDR & $0.099$ & $1.300$ & $1.692$\\
		& BFP & $0.104$ & $1.471$ & $1.897$\\
		& Spectral & $\bo{0.090}$ & $1.660$ & $2.478$\\
		\hline
	\end{tabular}
	\caption{Mean costs of GLL-based layouts ($\pm$ standard 
		error for SBM simulation experiment).
		The smallest quantity (within one standard error) in each 
		column for each experiment is bolded. 
		DGLL results using both a priori known groups (when available) and 
		groups learned by clustering are shown. 
		The regularizers present in DGLL lower both the centroid and 
		temporal costs at the expense of higher static cost.}
	\label{tab:Costs_DGLL}
\end{sidewaystable}

From Tables \ref{tab:Costs_DMDS} and \ref{tab:Costs_DGLL}, one can see that 
DMDS and DGLL have lower centroid and 
temporal costs than the baseline methods in all but one instance. 
Since DMDS and DGLL are the only methods to employ both grouping and temporal 
regularization, the results match up with what one might expect. 
The BFP method achieves a slightly lower centroid cost in the Newcomb 
experiment compared to DGLL but has a significantly higher temporal cost. 
The lower centroid and temporal costs for DMDS and DGLL are 
achieved by choosing node positions with a higher static cost. 
Notice also that DMDS requires significantly less iterations to converge 
than static MDS, which employs no regularization at all, and slightly 
less than stabilized MDS, which employs only temporal regularization. 
This is an added benefit of using both regularizers. 
The results for each experiment will be discussed in greater detail in the 
following. 

\subsection{Stochastic block model}
\label{sec:SBM}
In this experiment, we generate simulated networks using a stochastic block 
model (SBM) \citep{Holland1983}. 
An SBM creates networks with $k$ groups, where nodes 
in a group are stochastically equivalent,  
i.e.~the probability of forming an edge between nodes $i$ and $j$ 
is dependent only on the groups to which $i$ and $j$ belong. 
An SBM is completely specified by the set of probabilities $\{p_{cd}; \,
c = 1, \ldots, k; \, d = c, c+1, \ldots, k\}$, which represent the probability 
of forming an edge between any particular node in group $c$ and any particular 
node in group $d$. 

We generate $20$ independent samples from a $30$-node $4$-group SBM 
with parameters $p_{ii} = 0.6$ and $p_{ij} = 0.2$ for all $i \neq j$. 
Each sample corresponds to a graph snapshot at a single time step. 
The group memberships are randomly assigned at the initial time step 
and remain unchanged up to $t=9$. 
At $t=10$, $1/4$ of the nodes are randomly re-assigned to different 
groups to simulate a change in the network structure. 
The group memberships are then held constant until the last time step. 
We create layouts of the network using parameters $\alpha = \beta = 1$. 

\begin{figure}[t]
	\centering
	\subfloat{\includegraphics[width=2.4in]{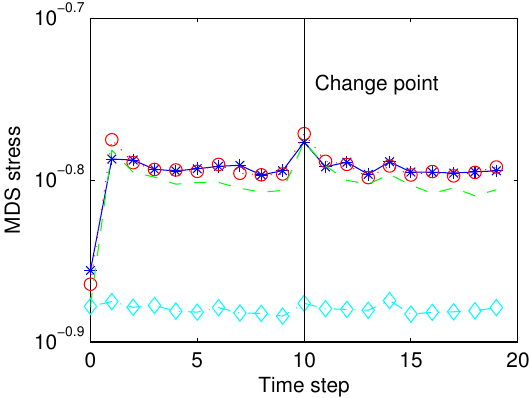}} 
	\quad
	\subfloat{\includegraphics[width=2.4in]{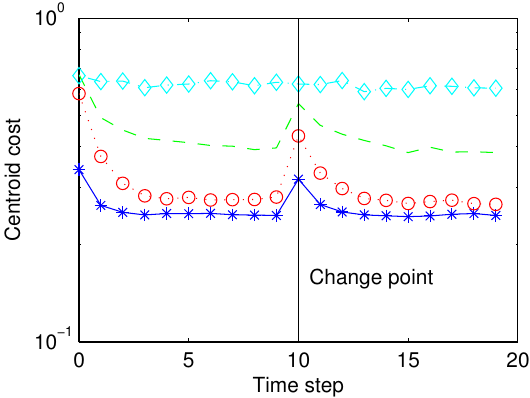}}\\
	\includegraphics[width=3.84in]{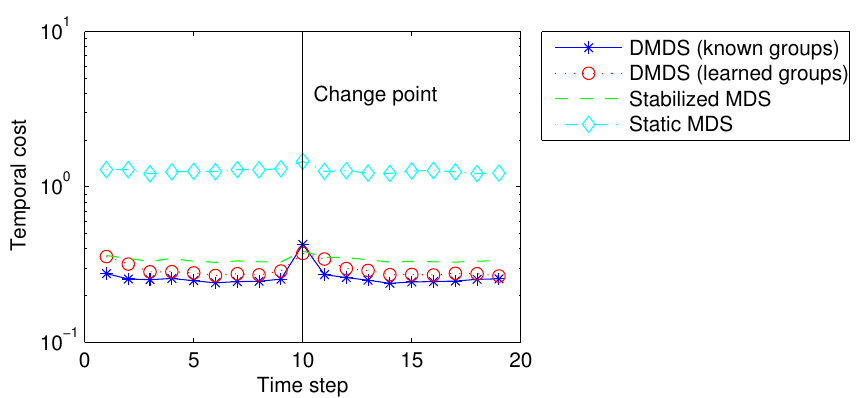}
	\caption{Costs of MDS-based layouts in the SBM experiment at 
		each time step. 
		The DMDS layouts have the lowest centroid and temporal costs but 
		also the highest MDS stress due to the regularizers.}
	\label{fig:SBM_costs_DMDS}
\end{figure}

\begin{figure}[t]
	\centering
	\subfloat{\includegraphics[width=2.4in]{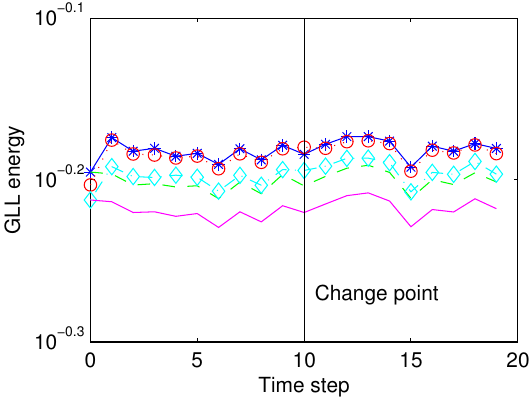}}
	\quad
	\subfloat{\includegraphics[width=2.4in]{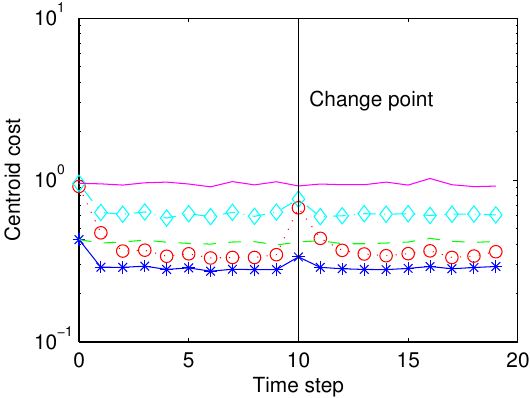}}\\
	\includegraphics[width=3.84in]{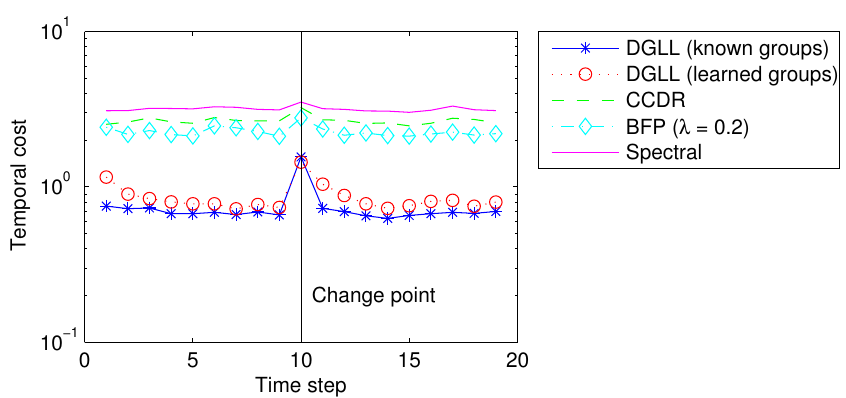}
	\caption{Costs of GLL-based layouts in the SBM experiment at 
		each time step.
		The DGLL layouts have the lowest centroid and 
		temporal costs but also the highest GLL energy due to the 
		regularizers.}
	\label{fig:SBM_costs_DGLL}
\end{figure}

In Fig.~\ref{fig:SBM_costs_DMDS}, we plot the variation over time of 
the static, centroid, and temporal costs of the MDS-based methods. 
The costs are averaged over $100$ simulation runs. 
The static cost is higher for the regularized layouts than for 
the static MDS layout. 
The grouping regularization in DMDS results in lower centroid cost as expected. 
When the groups are learned by clustering, the centroid cost is slightly 
higher than with the known groups, 
but still much lower than that of stabilized MDS and static MDS. 
Although stabilized MDS has only temporal regularization, notice that it 
also has a lower centroid cost than static MDS. 
This is because the SBM parameters are held constant from time steps $0$ to 
$9$ and from time steps $10$ to $19$, so that the group structure can be 
partially revealed by temporal regularization alone once enough time samples 
have been collected. 
The temporal regularization of both DMDS and stabilized MDS results in a 
significantly lower temporal cost than static MDS. 
The grouping regularization of DMDS also decreases the temporal cost slightly 
compared to stabilized MDS. 
An added benefit of the regularization in DMDS is the significant reduction 
in the number of iterations required for the MDS algorithm to converge, as 
shown in Table \ref{tab:Costs_DMDS}. 
On average, DMDS required less than half as many iterations as static MDS, and 
slightly less than stabilized MDS.

In Fig.~\ref{fig:SBM_costs_DGLL}, we plot the variation over time of 
the static, centroid, and temporal costs of the GLL-based methods. 
Similar to the MDS-based methods, the static cost is higher for the 
regularized layouts, but the centroid and temporal costs are much lower. 
Only DGLL is able to generate layouts with low temporal 
cost due to the temporal regularization. 
The grouping regularization in CCDR reduces the centroid cost but only 
slightly improves the temporal cost. 
The BFP method, which combines Laplacian matrices from two time steps, 
performs better than the standard spectral method both in centroid and 
temporal cost, but is worse than DGLL in both. 

Notice from Figs.~\ref{fig:SBM_costs_DMDS} and \ref{fig:SBM_costs_DGLL} 
that the centroid and temporal costs of DMDS and DGLL increase at $t=10$, 
reflecting the presence of the change in network structure. 
Such an increase is beneficial for two reasons. 
First, it indicates that the grouping and temporal penalties are not so 
strong that they mask changes in network structure, which would be 
undesirable. 
Second, it suggests that the centroid and temporal costs can be used to 
filter the sequence of networks for important events, such as change points, 
which can be useful for exploratory analysis of dynamic networks over 
long periods of time. 

\begin{figure}[t]
	\centering
	\includegraphics[width=4.9in]{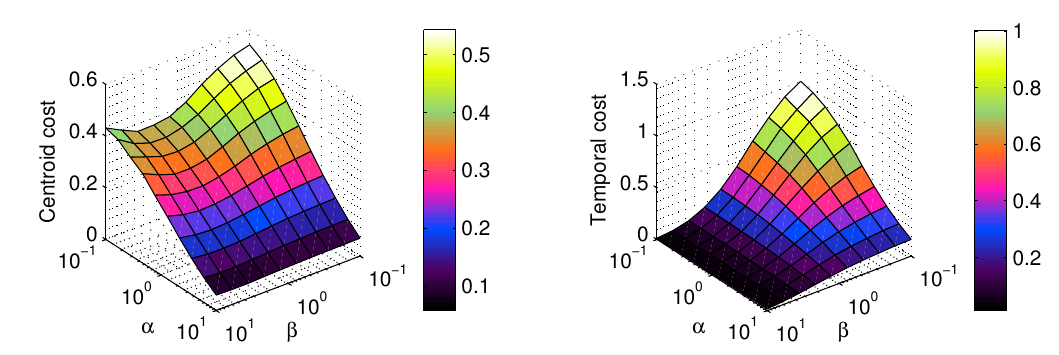}
	\caption{Mean centroid and temporal costs of DMDS layouts in the SBM 
		experiment as functions of $\alpha$ and $\beta$. 
		The centroid and temporal costs decrease as $\alpha$ and $\beta$ 
		are increased, respectively; however, $\alpha$ also affects the 
		temporal cost, and $\beta$ also affects the centroid cost.}
	\label{fig:SBM_param_DMDS}
\end{figure}

\begin{figure}[t]
	\centering
	\includegraphics[width=4.9in]{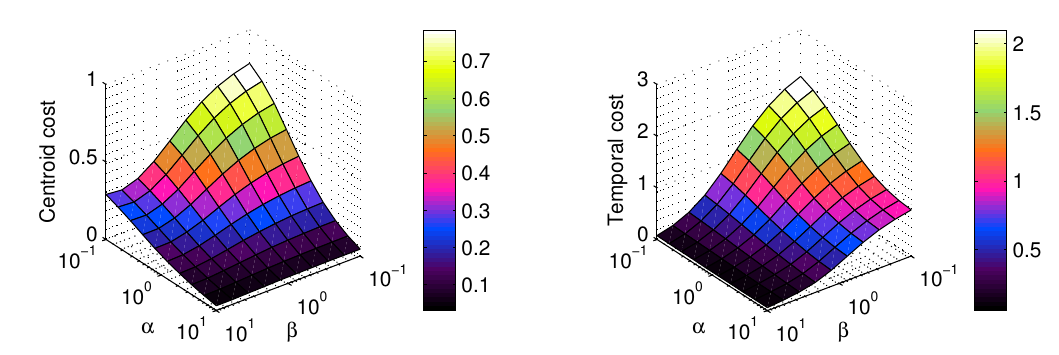}
	\caption{Mean centroid and temporal costs of DGLL layouts in the SBM 
		experiment as functions of $\alpha$ and $\beta$.
		The behavior of both costs as functions of 
		$\alpha$ and $\beta$ is similar to their behavior in DMDS.}
	\label{fig:SBM_param_DGLL}
\end{figure}

We demonstrate the effect of varying the regularization parameters in 
DMDS and DGLL, respectively, in Figs.~\ref{fig:SBM_param_DMDS} and 
\ref{fig:SBM_param_DGLL}. 
We generate layouts using $10$ choices each of $\alpha$ and $\beta$, 
uniformly distributed on a logarithmic scale between $0.1$ and $10$. 
The observations are similar for both DMDS and DGLL. 
As expected, the temporal cost decreases for increasing $\beta$. 
For low values of $\beta$, increasing $\alpha$ also 
decreases the temporal cost. 
This is a sensible result because nodes can move significantly over time 
but must remain close to the group representative, which lowers the 
temporal cost. 
The result is slightly different when it comes to the centroid cost. 
As expected, increasing $\alpha$ decreases centroid cost. 
For low values of $\alpha$, increasing $\beta$ also decreases centroid 
cost to a point, but a very high $\beta$ may actually increase centroid 
cost, especially in DMDS. 
This is also a sensible result because a very high 
$\beta$ places too much weight on the initial time step and prevents nodes 
from moving towards their group representative at future time steps. 

From this experiment we can see that there is a coupled effect between 
grouping and temporal regularization, and that using both regularizers 
can often result in both lower centroid and temporal costs. 
Indeed this is the case in all of the experiments in this paper, as shown 
in Tables \ref{tab:Costs_DMDS} and \ref{tab:Costs_DGLL}. 
However, it is important to note that this is not always true. 
For example, if a node changes group between two time steps, then the two 
penalties can oppose each other, with the temporal penalty 
attempting to pull the node towards its previous position and the grouping 
penalty attempting to pull the node towards its current representative, which 
could be quite far from the node's previous position. 
This is another reason for the increase in both centroid and temporal 
costs at $t=10$, when the group structure is altered, in 
Figs.~\ref{fig:SBM_costs_DMDS} and \ref{fig:SBM_costs_DGLL}.

\subsection{Newcomb's fraternity}
\label{sec:Newcomb}
This data set was collected by Nordlie and Newcomb \citep{Nordlie1958, 
Newcomb1961} as part of an experiment on interpersonal relations. 
It has been examined in several previous studies including 
\citet{Moody2005} and \citet{Bender-deMoll2006}. 
$17$ incoming male transfer students at the University of Michigan were 
housed together in fraternity housing. 
Each week, the participants ranked their preference of each of the other 
individuals in the house, in private, from $1$ to $16$. 
Data was collected over $15$ weeks in a semester, with one week of data 
missing during week $9$, corresponding to Fall break. 

We process the rank data in the same manner as 
\citet{Moody2005} and \citet{Bender-deMoll2006}. 
Graph snapshots are created by connecting each participant to his 
$4$ most preferred students with weights from $4$ decreasing to $1$ 
corresponding to the most preferred to the $4$th most preferred student. 
The graph is converted to an undirected graph by taking the edge weight 
between $i$ and $j$ to be the larger of the directed edge weights. 
The weights are converted into dissimilarities for the MDS-based methods 
by dividing each similarity 
weight by the maximum similarity of $4$. 
No group information is known a priori, so the group structure is learned 
using the AFFECT clustering algorithm. 

\begin{figure}[t]
	\centering
	\includegraphics[width=3.8in]{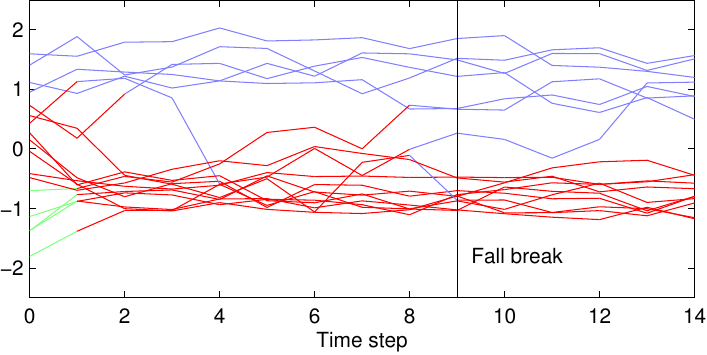}
	\caption{Time plots of $1$-D DGLL layouts of Newcomb's 
		fraternity, colored by learned groups. 
		Node positions in the layout are relatively stable over time 
		unless nodes are changing group.}
	\label{fig:Newcomb_1D_DGLL}
\end{figure}

In Fig.~\ref{fig:Newcomb_1D_DGLL}, we show a time plot of $1$-D layouts 
created using DGLL, 
where the color of a line segment between time steps $t$ and $t+1$ 
denotes the group membership of the node at time step $t$, 
and the location of the endpoints correspond to the node's position in 
the layouts at time steps $t$ and $t+1$. 
The regularization parameters are chosen to be $\alpha = \beta = 1$. 
While a $1$-D layout does a poor job of conveying the topology of 
the network, some temporal trends can be seen. 
For example, two mostly stable groups form after several weeks, but three 
students (numbers 10, 14, and 15) switch from the red to the blue group 
around Fall break. 
Student 15 reverts back to the red group after the break. 
Students 10 and 15 were found by \citet{Moody2005} to continuously 
change their preferences throughout the observation period. 
A similar observation can be made from the $1$-D DGLL layouts in 
Fig.~\ref{fig:Newcomb_1D_DGLL}, where students 10 and 15 have the two largest 
cumulative movements\footnote{The cumulative movement of node $i$ is 
measured by $\displaystyle\sum_{t=2}^{15} (\vec{x}_{(i)}[t] 
- \vec{x}_{(i)}[t-1])^2$.} of 
all the students ($4.51$ and $2.51$, respectively, compared to the mean over 
all students of $1.16$).

\begin{figure}[tp]
	\centering
	\includegraphics[width=4.9in]{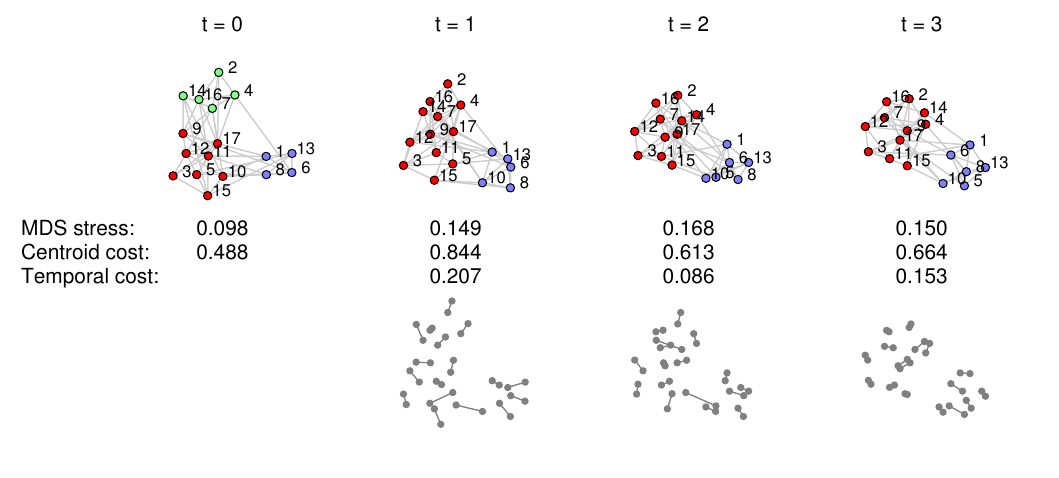}
	\caption{Layouts of Newcomb's fraternity at four time steps (top row) 
		generated using proposed DMDS algorithm and 
		node movements between layouts (bottom row). 
		The groups remain well-separated.}
	\label{fig:Newcomb_DMDS}
\end{figure}

\begin{figure}[tp]
	\centering
	\includegraphics[width=4.9in]{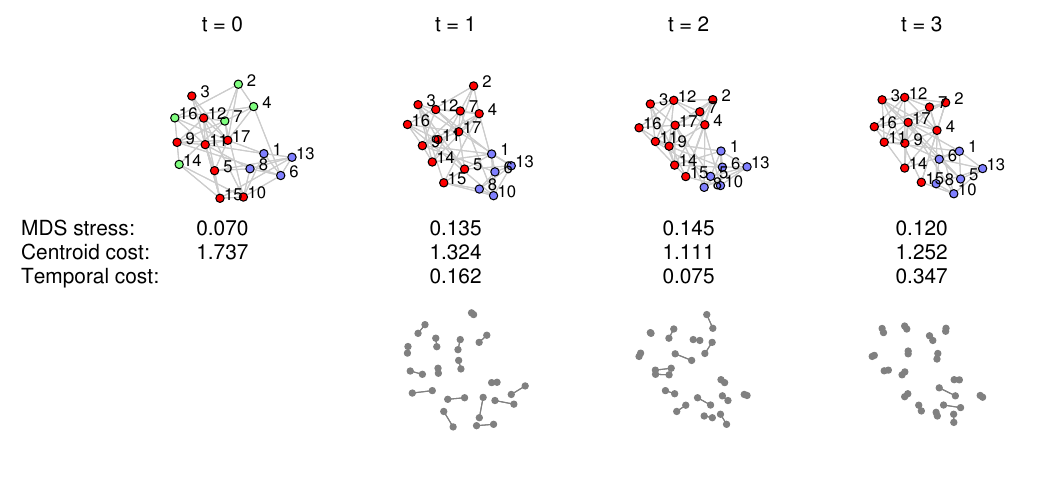}
	\caption{Layouts of Newcomb's fraternity at four time steps (top row) 
		using stabilized MDS and node movements between 
		layouts (bottom row). 
		The groups are not as well-separated as in the DMDS layouts.}
	\label{fig:Newcomb_Visone}
\end{figure}

\begin{figure}[tp]
	\centering
	\includegraphics[width=4.9in]{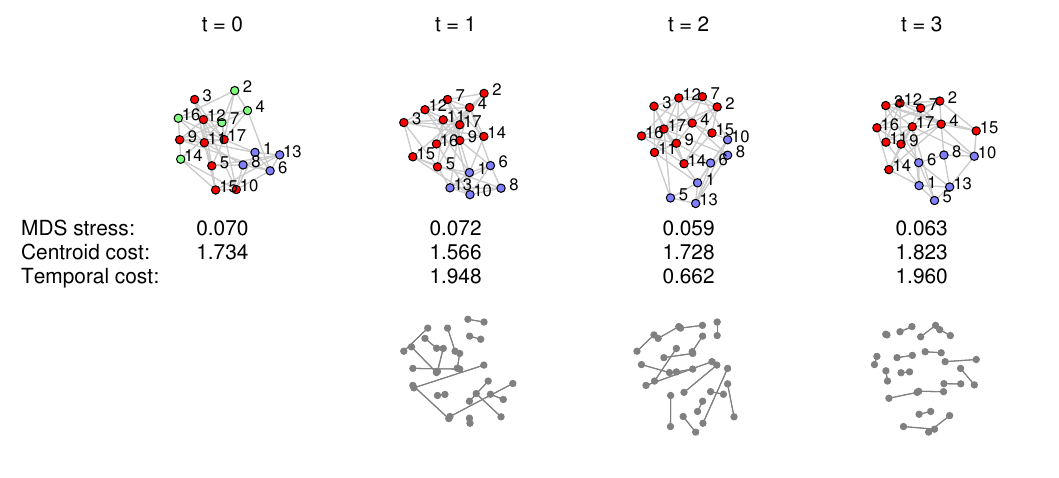}
	\caption{Layouts of Newcomb's fraternity at four time steps (top row) 
		using static MDS and node movements between layouts 
		(bottom row). 
		There is excessive node movement, and the groups are not as 
		well-separated as in the DMDS layouts.}
	\label{fig:Newcomb_SoNIA}
\end{figure}

In Figs.~\ref{fig:Newcomb_DMDS}-\ref{fig:Newcomb_SoNIA}, we present a 
comparison of the first four snapshots from the layouts created using 
DMDS, stabilized MDS, and static MDS, respectively. 
In both figures, the top row corresponds to the layouts, and the bottom 
row illustrates the movement of each node over time. 
In the plots on the bottom row, each node is drawn twice: once at its 
current position at time $t$ and once at its previous position at time 
$t-1$. 
An edge connects these two positions; the length of the edge indicates 
how far a node has moved between time steps $t-1$ and $t$. 

At $t=0$, the red and green groups are mixed together in the stabilized MDS 
and static MDS layouts, while they are easily distinguished in the DMDS 
layout due to the grouping regularization. 
Furthermore, the node movements over time in the static MDS layouts are much 
more extreme than in the DMDS layouts. 
The excessive node movement is even more visible by comparing the 
DMDS and static MDS animations on the supporting website \citep{XuSuppWeb}. 
The excessive node movement is reflected in the substantially 
higher temporal cost of 
static MDS compared to DMDS, as shown in Table \ref{tab:Costs_DMDS}, due 
to the lack of temporal regularization. 
DMDS also has lower mean centroid and temporal costs compared to stabilized 
MDS, although the improvement in temporal cost is smaller than compared to 
static MDS because stabilized MDS also employs temporal regularization.  
Finally, the mean number of iterations required for the static MDS 
layout to converge is almost four times that of DMDS, so DMDS presents 
significant computational savings in addition to better preservation of 
the mental map.

For the GLL-based layouts, which can be found on the supporting website 
\citep{XuSuppWeb}, there is not much difference in terms of the 
centroid cost. 
Notice from Table \ref{tab:Costs_DGLL}, that the mean centroid cost differs 
by only $4\%$ between the best (BFP) and worst (spectral). 
This is not surprising, as the groups are learned using evolutionary 
spectral clustering, which is closely related to GLL. 
However, DGLL achieves significantly lower temporal cost than the baseline 
methods due to the temporal regularization. 
The temporal smoothing of the Laplacian for BFP helps to lower the temporal 
cost slightly compared to the CCDR and static spectral layouts, 
but it does not explicitly encourage stability like the temporal 
regularization in DGLL.

\subsection{MIT Reality Mining}
\label{sec:Reality}
The MIT Reality Mining data set \citep{EaglePNAS2009} was collected as 
part of an experiment on inferring social networks by using cell 
phones as sensors. 
$94$ students and staff at MIT were given access to smart phones that 
were monitored over two semesters. 
The phones were equipped with Bluetooth sensors, and each phone recorded the  
Media Access Control addresses of nearby Bluetooth devices at 
five-minute intervals. 
Using this proximity data, we construct 
a sequence of graph snapshots where each participant is connected to the 
$5$ participants he or she was in highest proximity to during a time step. 
We divide the data into 
time steps of one week, resulting in $46$ time steps between August 
2004 and June 2005. 
From the MIT academic calendar \citep{MITCal200405}, we 
know the dates of important events such as the beginning and end of school 
terms. 
We also know 
that $26$ of the participants were incoming students at the university's 
business school, while the rest were colleagues working in the same building. 
These affiliations are used as the known groups. 

\begin{figure}[t]
	\centering
	\includegraphics[width=4.9in]{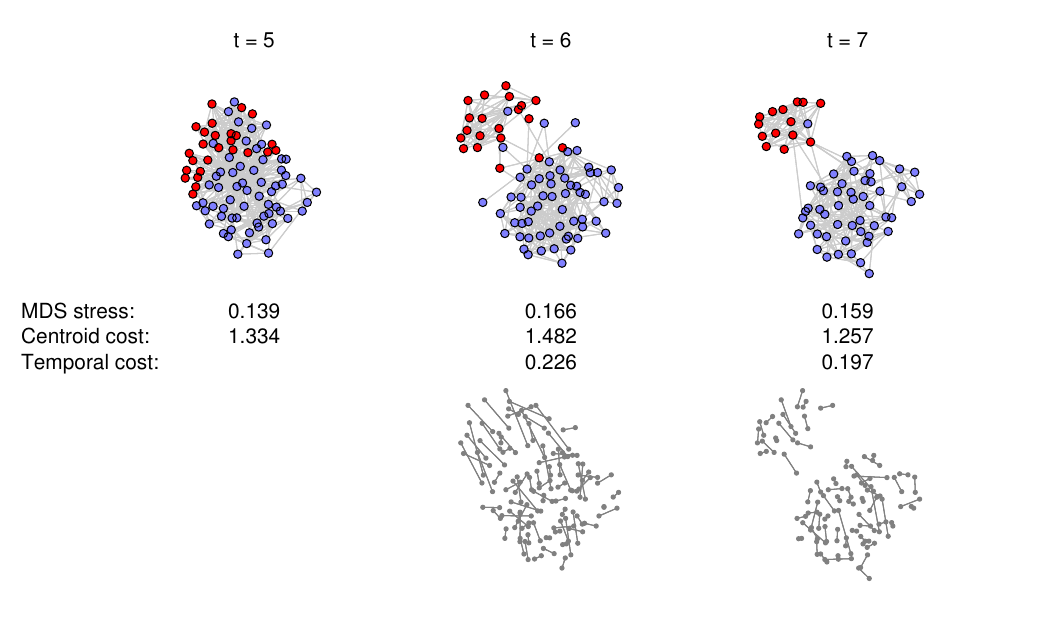}
	\caption{DMDS layouts of MIT Reality Mining data at four time steps 
		using the known groups (top row) and node movements between layouts 
		(bottom row). 
		Blue nodes denote colleagues working in the same building, 
		and red nodes denote incoming students. 
		The incoming students separate from the others after the first 
		week of classes ($t=5$).}
	\label{fig:MIT_DMDS}
\end{figure}

The DMDS layouts at three time steps computed using the using the known 
groups with $\alpha = 1, \beta = 3$ are shown in Fig.~\ref{fig:MIT_DMDS}. 
A higher value of $\beta$ is chosen compared to the previous experiments in 
order to create more stable layouts due to the higher number of nodes. 
Node labels are not displayed to reduce clutter in the figure. 
We encourage readers to view the animation on the supporting website 
\citep{XuSuppWeb} 
to get a better idea of the temporal evolution of this network. 
$t=5$ corresponds to the first week of classes. 
Notice that the two groups are slightly overlapped at this time step. 
As time progresses, the group of incoming students separates quite clearly 
from the colleagues working in the same building. 
This result suggests that the incoming students are spending more time in 
proximity with each other than with the remaining participants, which 
one would expect as the students gain familiarity with each other as the 
semester unfolds. 

\begin{figure}[t]
	\centering
	\includegraphics[width=4.9in]{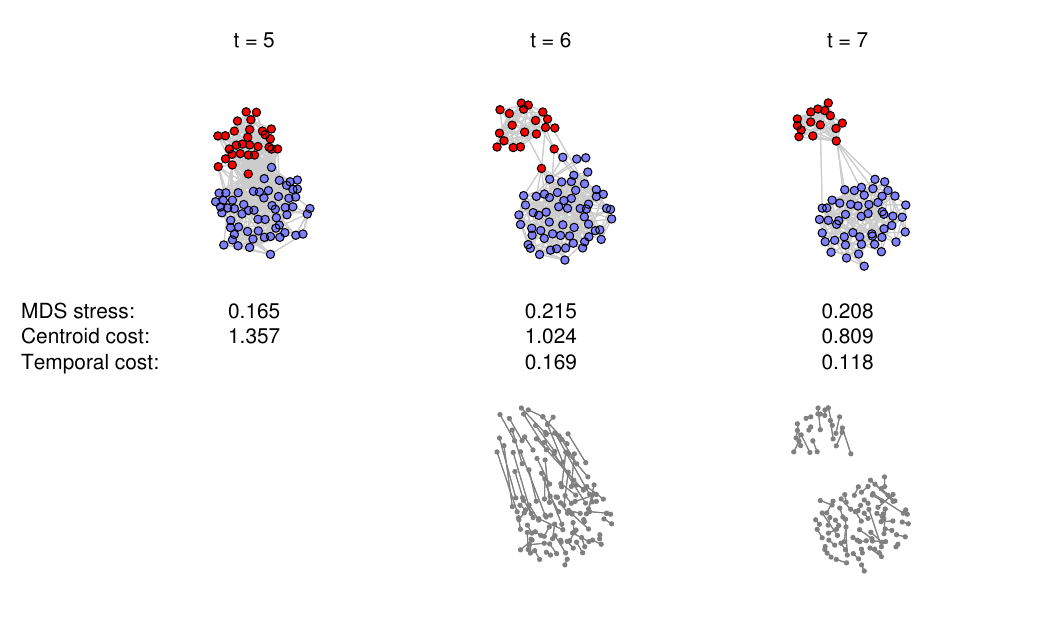}
	\caption{DMDS layouts of MIT Reality Mining data at four time steps 
		with $\alpha=5, \beta=3$ using groups learned by clustering 
		(top row) and node movements between layouts (bottom row). 
		Colors correspond to learned groups. 
		There is a lot of node movement between groups but very little 
		movement within groups, resulting in high MDS stress.}
	\label{fig:MIT_high_alpha}
\end{figure}

\begin{figure}[t]
	\centering
	\includegraphics[width=4.9in]{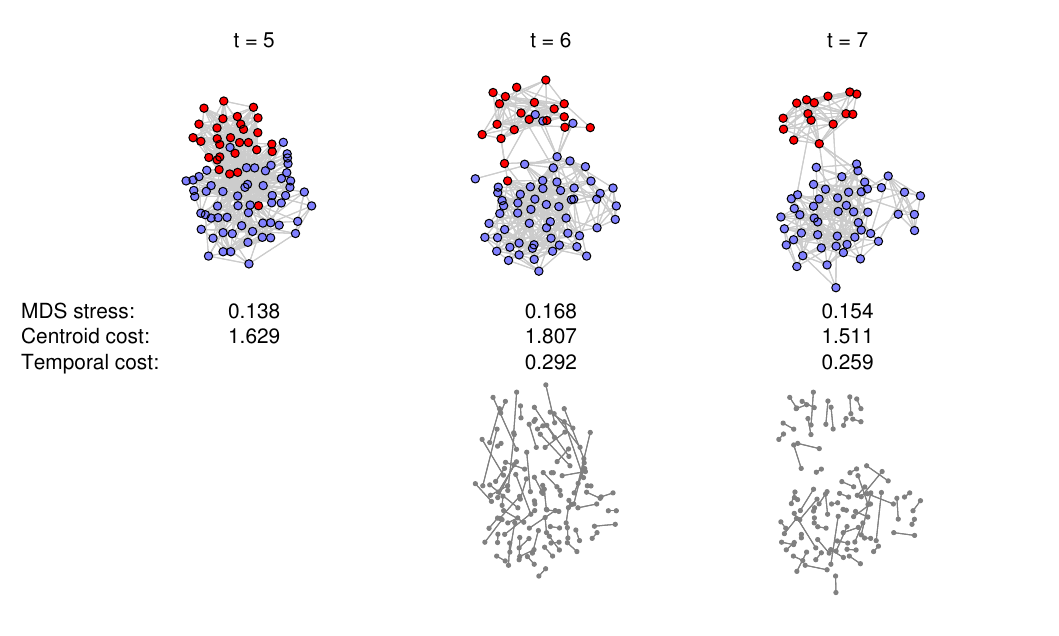}
	\caption{DMDS layouts of MIT Reality Mining data at four time steps 
		with $\alpha=1/5, \beta=3$ using groups learned by clustering  
		(top row) and node movements between layouts (bottom row). 
		Colors correspond to learned groups. 
		There is more movement within groups, resulting in lower MDS 
		stress, but it is more difficult to identify movement between 
		groups.}
	\label{fig:MIT_low_alpha}
\end{figure}

The same observation can be made from the DMDS layouts computed using 
the groups learned by the AFFECT clustering algorithm. 
Initially at $t=5$, the separation between groups is not clear so many 
nodes are not correctly classified, but at subsequent time steps when the 
separation is clearer, almost all of the nodes are correctly classified. 
Between time steps $5$ and $6$ many nodes switch groups. 
This can be seen in Fig.~\ref{fig:MIT_high_alpha}, where the colors 
correspond to the learned groups rather than the known groups. 
These layouts are created using $\alpha=5,\beta=3$; the high value of 
$\alpha$ emphasizes the node movements between groups while sacrificing 
the quality of movements within groups, as discussed in Section 
\ref{sec:Discussion}. 
Notice that the groups are very compact and well-separated, so that nodes 
switching from one group to another have large movements between layouts. 
Compare these layouts to those shown in Fig.~\ref{fig:MIT_low_alpha}, 
which are created using $\alpha=1/5,\beta=3$. 
The lower value of $\alpha$ better shows movements within 
groups, but the large changes in groups between time steps $5$ and $6$ 
are not as obvious as in Fig.~\ref{fig:MIT_high_alpha}. 
Both layouts are useful and provide different insights into the 
network dynamics; however, the observation of the incoming students 
separating from the other participants is evident in both visualizations. 

The benefits of using both regularizers can be seen once again from the 
statistics in Tables \ref{tab:Costs_DMDS} and \ref{tab:Costs_DGLL}. 
With known groups, the DMDS and DGLL layouts have lower mean 
centroid and temporal costs compared to all baseline layouts. 
The DMDS and DGLL layouts using groups learned by 
clustering still have lower temporal cost than the baseline methods, and only 
CCDR, which uses the known groups, achieves lower centroid cost than DGLL 
with learned groups. 
The DMDS algorithms also converge more quickly than both stabilized MDS and 
static MDS. 

\section{Conclusions}
In this paper we proposed a regularized graph layout framework for 
dynamic network visualization. 
The proposed framework incorporates both grouping and temporal regularization 
into graph layout in order to discourage nodes from deviating too far from 
other nodes in the same group and from their previous position, 
respectively. 
The layouts are generated in an on-line manner using only present and 
past data. 
We introduced two dynamic layout algorithms, DMDS and DGLL, which are 
regularized versions of their static counterparts. 
Multiple experiments demonstrate that the regularizers do indeed have the 
intended effects of lowering the centroid and temporal costs to better 
preserve the mental map.

An important area for future work concerns visualization of extremely 
large dynamic networks containing upwards of thousands of nodes. 
One issue is related to scalability; both DMDS and DGLL require $O(n^3)$ 
computational time and thus may not be applicable to extremely large networks. 
An even more significant issue involves the interpretation of visualizations 
of such large networks. 
Even when equipped with grouping and temporal regularization, 
layouts of extremely large networks may be confusing for a human to interpret 
so additional techniques may be necessary to deal with this challenge. 

\appendixtitleon
\appendixtitletocon
\begin{appendices}
\section{DGLL solution in 2-D}
\label{sec:DGLL_2D}
We derive the expressions for $\nabla f$, $g$, $H$, and $J$ in $2$-D. 
These vectors and matrices are computed at each iteration in the DGLL 
algorithm to solve \eqref{eq:DGLL_obj} using the interior-point 
algorithm \citep{Byrd1999} as discussed in Section \ref{sec:DGLL}. 
The constraints can be written as $g(\tilde{X}) = 0$ where
\begin{equation*}
	g(\tilde{X}) = 
	\begin{bmatrix}
		\tilde{\bo x}_1^T M \tilde{\bo x}_1 - \tr(\tilde{D}) \\
		\tilde{\bo x}_2^T M \tilde{\bo x}_2 - \tr(\tilde{D}) \\
		\tilde{\bo x}_2^T M \tilde{\bo x}_1
	\end{bmatrix}.
\end{equation*}
The gradient of the objective function is given by
\begin{equation*}
	\nabla f(\tilde{X}) = 
	\begin{bmatrix}
		(2\tilde{L} + 2\beta\tilde{E}) \tilde{\bo x}_1 
			- 2\beta\tilde{E}\tilde{\bo x}_1[t-1] \\
		(2\tilde{L} + 2\beta\tilde{E}) \tilde{\bo x}_2 
			- 2\beta\tilde{E}\tilde{\bo x}_2[t-1]
	\end{bmatrix}.
\end{equation*}
The Jacobian of the constraints is given by 
\begin{equation*}
	J(\tilde{X}) = 
	\begin{bmatrix}
		2\tilde{\bo x}_1^T M & \bo 0 \\
		\bo 0 & 2\tilde{\bo x}_2^T M \\
		\tilde{\bo x}_2^T M & \tilde{\bo x}_1^T M
	\end{bmatrix}.
\end{equation*}
Finally, the Hessian is obtained by 
\begin{align*}
	H(\tilde{X}, \bm \mu) &= \nabla^2 f(\tilde{X}) + \mu_1 \nabla^2 
		g_1(\tilde{X}) + \mu_2 \nabla^2 g_2(\tilde{X}) + \mu_3 \nabla^2 
		g_3(\tilde{X}) \\
	&= \begin{bmatrix}
		2\tilde{L} + 2\beta\tilde{E} + 2\mu_1 M & \mu_3 M \\
		\mu_3 M & 2\tilde{L} + 2\beta\tilde{E} + 2\mu_2 M
	\end{bmatrix}.
\end{align*}

\section{AFFECT evolutionary clustering algorithm}
\label{sec:AFFECT}
Evolutionary clustering 
algorithms are designed to cluster dynamic data where a set of objects is 
observed over multiple time steps. 
In the dynamic network setting, objects correspond to nodes, and observations 
correspond to graph adjacency matrices $W[t]$. 
The AFFECT (adaptive forgetting factor for evolutionary clustering and 
tracking) framework \citep{Xu2013} involves creating a \emph{smoothed 
adjacency matrix} at each time step and then performing ordinary static 
clustering on this matrix. 
The smoothed adjacency matrix is given by 
\begin{equation*}
	\hat\Psi[t] = \alpha[t] \hat{\Psi}[t-1] + (1-\alpha[t])W[t],
\end{equation*}
where $\alpha[t]$ is a forgetting factor that controls how quickly previous 
adjacency matrices are forgotten. 
We drop the time index for quantities at time $t$ for simplicity. 
The AFFECT framework adaptively estimates the 
optimal amount of smoothing to apply at each time step in order to minimize 
mean-squared error (MSE) in terms of the Frobenius norm $\E[\|\hat{\Psi} 
- \Psi\|_F^2]$, where $\Psi$ denotes the expected adjacency matrix $\E[W]$, 
which can be viewed as a matrix of unknown states characterizing the network 
structure at time $t$.
It is shown in \citet{Xu2013} that the optimal choice of $\alpha$ 
is given by
\begin{equation*}
	\alpha^* = \frac{\sum_{i=1}^n \sum_{j=1}^n \var\left(w_{ij}\right)}
		{\sum_{i=1}^n \sum_{j=1}^n \left\{\left(\hat\psi_{ij}[t-1] 
		- \psi_{ij}\right)^2 + \var\left(w_{ij}\right)\right\}}.
\end{equation*}
For real networks, $\psi_{ij}$ and $\var(w_{ij})$ are unknown so $\alpha^*$ 
cannot be computed. 

The AFFECT framework iteratively estimates the optimal forgetting factor 
and clusters nodes using a two-step procedure. 
Begin by initializing the clusters, for example, by using the clusters at the 
previous time step. 
$\alpha^*$ is estimated by replacing the unknown means and variances 
with sample means and variances computed over the clusters. 
Static clustering is then performed on the smoothed adjacency matrix 
$\hat{\Psi}$ to obtain cluster memberships. 
The estimate of $\alpha^*$ and the cluster memberships are then refined by 
iterating the two steps. 
In this paper, we use normalized cut spectral clustering \citep{Ng2001} 
as the static clustering algorithm. 
We refer interested readers to \citet{Xu2013} for additional details.
\end{appendices}

\section*{Acknowledgements}
This work was partially supported by the 
Office of Naval Research grant N00014-08-1-1065, the 
National Science Foundation grant CCF 0830490, and the 
Army Research Office grant W911NF-09-1-0310. 
Kevin Xu was partially supported by an award from the Natural 
Sciences and Engineering Research Council of Canada.

\bibliographystyle{abbrvnat}
\bibliography{library_s,full,references,Xu_visualization_DMKD_2012}

\end{document}